\documentclass[prd,floatfix, nofootinbib, preprint]{revtex4-1}		%
\usepackage[utf8]{inputenc}
\usepackage{amssymb}
\usepackage{amsmath}
\usepackage{bm, natbib}
\usepackage{braket}
\usepackage[mathscr]{euscript}
\usepackage{float}
\usepackage{graphicx}
\usepackage{subfig}
\usepackage{xcolor}
\usepackage{tikz}
\usetikzlibrary{positioning,arrows}
\usetikzlibrary{decorations.pathmorphing}
\usetikzlibrary{decorations.markings}
\usetikzlibrary{shapes.geometric}
\usepackage{times}
\usepackage{multirow}
\usepackage{slashed}

\tikzset{
    vector/.style={decorate, decoration={snake}, draw},
    provector/.style={decorate, decoration={snake,amplitude=2.5pt}, draw},
    antivector/.style={decorate, decoration={snake,amplitude=-2.5pt}, draw},
    fermion/.style={draw=black, postaction={decorate},decoration={markings,mark=at position .55 with {\arrow[draw=black]{>}}}},
    fermionbar/.style={draw=black, postaction={decorate},
                       decoration={markings,mark=at position .55 with {\arrow[draw=black]{<}}}},
    fermionnoarrow/.style={draw=black},
    gluon/.style={decorate, draw=black,decoration={coil,amplitude=4pt, segment length=5pt}},
    scalar/.style={dashed,draw=black, postaction={decorate},decoration={markings,mark=at position .55 with {\arrow[draw=black]{>}}}},
    scalarbar/.style={dashed,draw=black, postaction={decorate},decoration={markings,mark=at position .55 with {\arrow[draw=black]{<}}}},
    scalarnoarrow/.style={dashed,draw=black},
    electron/.style={draw=black, postaction={decorate},decoration={markings,mark=at position .55 with {\arrow[draw=black]{>}}}},
    bigvector/.style={decorate, decoration={snake,amplitude=4pt}, draw},
}
\tikzset{
    vector/.style={decorate, decoration={snake}, draw},
    provector/.style={decorate, decoration={snake,amplitude=2.5pt}, draw},
    antivector/.style={decorate, decoration={snake,amplitude=-2.5pt}, draw},
    fermion/.style={draw=black, postaction={decorate},decoration={markings,mark=at position .55 with {\arrow[draw=black]{>}}}},
    fermionbar/.style={draw=black, postaction={decorate},
                       decoration={markings,mark=at position .55 with {\arrow[draw=black]{<}}}},
    fermionnoarrow/.style={draw=black},
    gluon/.style={decorate, draw=black,decoration={coil,amplitude=4pt, segment length=5pt}},
    scalar/.style={dashed,draw=black, postaction={decorate},decoration={markings,mark=at position .55 with {\arrow[draw=black]{>}}}},
    scalarbar/.style={dashed,draw=black, postaction={decorate},decoration={markings,mark=at position .55 with {\arrow[draw=black]{<}}}},
    scalarnoarrow/.style={dashed,draw=black},
    electron/.style={draw=black, postaction={decorate},decoration={markings,mark=at position .55 with {\arrow[draw=black]{>}}}},
    bigvector/.style={decorate, decoration={snake,amplitude=4pt}, draw},
}
\begin{document}
\title{Searching for New physics in Charm Radiative decays}
\author{Aritra Biswas\footnote{aritrab@imsc.res.in}, Sanjoy Mandal\footnote{smandal@imsc.res.in}, Nita Sinha\footnote{nita@imsc.res.in}}
\affiliation{The Institute of Mathematical Sciences, C.I.T Campus, Taramani, Chennai 600 113, India} 
\affiliation{Homi Bhabha National Institute, BARC Training School Complex,
Anushakti Nagar, Mumbai 400094, India.}
\begin{abstract}
We show that for a heavy vector-like quark model with a down type isosinglet, branching ratio for $c\rightarrow u\gamma$ decay is enhanced by more than $\mathcal{O}(10^2)$ 
as compared to that in the Standard model when QCD corrections to next-to-leading order are incorporated. %If one considers a \tcr{model with} left-right symmetry along with \tcr{a} heavy vector-like fermion, for specific values of the parameters of the model enhancement of this order can be achieved at the bare (QCD uncorrected ) level itself. We also point out that a measurement of the photon polarization could be possibly used for identifying such new physics.
In a left-right symmetric model (LRSM) along with a heavy vector-like fermion, enhancement of this order can be achieved at the bare (QCD uncorrected) level itself. 
 % in analogy to $b\rightarrow s\gamma$ decays.
 %\tcr{Inspite of the presence of long distance effects,}  we find that there is a large region within the allowed parameter space of the left-right symmetric model as well as the model with vector-like down type isosinglet quark with additional left-right symmetry, %for such scenarios 
We propose that a measurement of the photon polarization could be used to signal the presence of such new physics inspite of the large long distance effects.  We find that there is a large region within the allowed parameter
space of model with vector-like quark with additional left-right symmetry,
 where depending on the exact size of the long distance contribution, the photon polarization can be dominantly right-handed.
\end{abstract}
\maketitle
\section{Introduction}
The direct search for physics beyond the Standard Model(SM) has been unsuccessful thus far. There have been anomalies in some of the observables in the flavour sector, with deviations from the SM predictions at the level of few 
sigma~\cite{Ligeti}. In fact, the presence of New Physics(NP) at possibly high scales may very well be deduced only from precision measurements of some of the rare meson decays. The absence of Flavour changing neutral currents (FCNCs) at the tree level allows 
the possibility of virtual new physics particles to be present in the loop level diagrams that contribute to these processes. Detailed study of rare charm decays was performed in Ref.~\cite{pakvasa_2002}.
Predictions for these decays in various extensions of the SM including extensions of the Higgs, gauge and fermion sectors were obtained.
Rare charm decays were also recently studied in Refs.~\cite{fajfer_rare,petrov_rare,Boer}. The focus of this study will be the radiative decays of charmed mesons. While both the inclusive and 
exclusive radiative $B$ meson decays have been extensively discussed in the literature, less attention has been paid to the $D$ meson radiative decays as their branching ratios are expected to be much smaller due to the almost 
complete GIM suppression.

Moreover, charm radiative decays will be dominated by long distance contributions, which can hide the presence of new physics particles that may appear in the loop of the short distance penguin contributions. Nevertheless, in Ref.~\cite{FajferRatio} it was pointed out that a measurement of the difference in the rates of the exclusive modes, $D^0\to\rho\gamma$ and $D^0\to\omega\gamma$ in which the long distance effects are expected to cancel, would indicate short distance new physics if the data reveals a difference of rates which is more than $30\% $. But in general, due to the large uncertainties in the long 
distance contributions, any definite conclusion regarding NP will not be feasible from a measurement of the radiative decay rates for the inclusive $c\to u\gamma$ case nor for any individual exclusive channel, unless the NP short distance contribution is larger than that from the long distance effects. In fact, the possibility of enhancement above the otherwise dominant long distance effects, in presence of a fourth generation model with large mixing angles of the $b^\prime$ quark, $U_{ub^\prime}U_{ub^\prime}$ had been pointed out in Ref.~\cite{BabuHeLiPakvasa}. Fourth generation models are now inconsistent with the LHC data, however, models with vector-like charge $-1/3$ quarks, for which the authors of Ref.~\cite{BabuHeLiPakvasa} claimed that their results were also applicable, are still viable. In fact, in the last couple of years many detailed studies of the phenomenology of vector-like quarks and constraints from the flavour sector have been performed~\cite{Cacciapaglia2011,Botella2013,Ishiwata2015,Bobeth2016,uma}. 
 
Apart from the enhancement in the decay rate, which will be subject to the relative size of the short distance and long distance effects, NP could also be searched through a measurement of the polarization of the photon produced in the decay. The SM has  a robust prediction regarding the photon 
polarization in $c\to u\gamma$ decays and hence, a measurement of the photon polarization can pin down the presence of NP. This had been earlier pointed out for the case of $B$ radiative decays in Refs.~\cite{Atwood, Gronau}. In the SM, the photons %emitted 
from the short distance (SD) penguin contribution in the $c\to u\gamma$ decays will be mostly left handed up to corrections of 
$\cal{O}$($m_u/m_c$). This dominance of left handed polarization can get masked in the presence of long distance (LD) effects. However, the fraction of the right polarized photons will vary in different models and may possibly even allow one to 
distinguish between different models of NP. We explore the effects of the presence of a down-type isosinglet vector-like quark model on the $c\to u\gamma$ decay rate, as well as on the photon polarization for this model with an additional left-right symmetry. The decay rate evaluation requires an estimation of both the SD as well as LD components, which are described in the next section. In Sec.~\ref{VLQ} some details of the down type isosinglet vector-like quark model are discussed, including the modifications to the Wilson coefficients in its presence. Sec.~\ref{LRSM} contains a short discussion on LRSM and the results for the bare level SD contributions to the amplitudes for the emission of the left and right handed photons in this model. Sec.~\ref{IVA} gives our results for the branching ratios (BR's) in the SM and in the different NP models. In Sec.~\ref{IVB} we present our analysis of the polarization function in the LRSM and the LRSM with vector-like quark. Finally in Sec.~\ref{V}, we conclude.
    
\section{The long and short distance contributions within the Standard Model}
The long distance contributions being non-perturbative in nature are hard to estimate. The long distance contributions can come either from pole diagrams or vector meson dominance (VMD) diagrams. At the quark level, the pole contribution corresponds to the annihilation diagrams 
$c\overline{q}_{1}\rightarrow q_{2}\overline{q}_{3}$ with a photon attached to any of the four quark lines. They are actually a subset of a more general class of long distance contributions, which include two-particle 
intermediate states and extends up to all higher n-particle intermediate states. Phenomenologically however, the single-particle or pole terms are the most accessible. The underlying quark processes in the VMD contributions are 
$c\rightarrow q_{1}\overline{q}_{2}q$, followed by $\overline{q}_{2}q\rightarrow\gamma$. All these long distance effects are rather hard to calculate from first principles but
can be estimated in models. Hence, it is important that the observables chosen for uncovering short distance NP,
have different values from the SM case, even in the presence of the large long distance contributions. We provide an updated estimate for the long distance amplitudes and branching ratios for charm decays following the methods of Ref.~\cite{Burdman} in appendix~\ref{LDBR}.  

\begin{figure}
\begin{minipage}{0.49\textwidth}
\begin{tikzpicture}[line width=1 pt, scale=1.1]
\draw (-5,0) -- (-1.5,0);
\draw[vector] (-4,0) arc (180:.5:.8);
\draw[vector] (-2.8,0.8)--(-1.7,1.8);
\node at (-4.1,0.8) {\small W};
\node at (-2.3,0.8) {\small W};
\node at (-5.2,0) {c};
\node at (-1.3,0) {u};
\node at (-2.1,1.8) {$\gamma$};
\node at (-3.2,-0.2) {d,s,b};
\end{tikzpicture}
\end{minipage}
\begin{minipage}{0.49\textwidth}
\begin{tikzpicture}[line width=1 pt, scale=1.1]
\draw (-5,0) -- (-1.5,0);
\draw[vector] (-4,0) arc (-180:.5:.8);
\draw[vector] (-3.2,0)--(-1.7,1);
\node at (-3.2,-1.2) {\small W};
\node at (-5.2,0) {c};
\node at (-1.3,0) {u};
\node at (-2.1,1.1) {$\gamma$};
\node at (-3.2,-0.2) {d,s,b};
\end{tikzpicture}
\end{minipage}
\caption{The Feynman diagrams for the process $c\to u\gamma$.}
\label{ctougamma fig}
\end{figure}
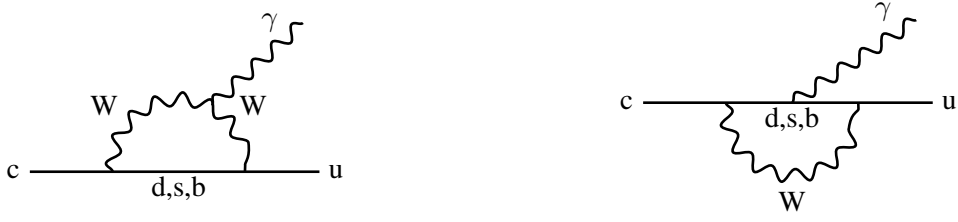

The amplitude for the flavour changing radiative transitions were first evaluated by Inami and Lim~\cite{InamiLim}. As pointed out in Ref.~\cite{Kim}, those formulaes need to be appropriately modified for the case of $c\to u\gamma$  decay. 
 The SM Lagrangian for the processes $c\to u\gamma$, which arises at the loop level as shown in
 	Fig.\ref{ctougamma fig}  is given by,
 	\begin{align}\label{SM_Lag}
 	\mathcal{L}_{\text{int}}=-\frac{4G_{F}}{\sqrt{2}}A^{SM}\frac{e}{16\pi^{2}}m_{c}\left(\bar{u}\sigma_{\mu\nu}P_{R}c\right)F^{\mu\nu},
 	\end{align}
 	where the mass of the final quark $u$ has been neglected and $P_{R}=\frac{1+\gamma_5}{2}$. The coefficient $A^{SM}$ is a function of the internal quark masses and the (QCD uncorrected) contribution to the amplitude $A^{SM}$ is given by,
 	\begin{align}\label{Asm}
 	A^{SM}&=\sum_{p=1,2}Q_p\left[ V_{cb}^{*}V_{ub}G_p(r_{b})+V_{cs}^{*}V_{us}G_p(r_{s})+V_{cd}^{*}V_{ud}G_p(r_{d})\right]\nonumber\\
 	&=\sum_{p=1,2}Q_p\sum_{q=d, s, b}V_{ci}^{*}V_{ui}G_p(r_{q}),
 	\end{align}
        where %$F(r)=G_{1}(r)-\frac{1}{3}G_{2}(r)$ with
 	 $r_q=\frac{m_q^2}{M_{W}^2}$ with $m_q$ ($q=d$, $s$, $b$) being the masses of the down-type quarks running in the fermion loop in the penguin diagrams. The functions $G_p, p=1,2$ defined in~\cite{Kim} are given in appendix~\ref{G}. $Q_1$ and $Q_2$ are the charges of the $W$ boson emitted from the initial quark in the loop diagram, and that of the internal quark running in the loop, respectively.
 	The inclusive decay rate for a $c\rightarrow u\gamma$ process within the SM is given by,
 	\begin{align}
 	\Gamma^{0}_{c\to u\gamma}=\frac{\alpha G_{F}^{2}}{128\pi^{4}}m_{c}^{5}|A^{SM}|^{2}.
 	\end{align}
 	This results in the following inclusive BR for the $c\rightarrow u\gamma$ process,
 	\begin{align}
 	BR(c\to u\gamma)=\frac{3}{4}\frac{\alpha}{\pi}\frac{|A^{SM}|^{2}}{|V_{cs}|^{2}I(\frac{m_{s}^{2}}{m_{c}^{2}})+|V_{cd}|^{2}I(\frac{m_{d}^{2}}{m_{c}^{2}})}
 	BR(D^{+}\to Xe^{+}\nu_{e}),
 	\end{align}
 	which is normalized with respect to the inclusive semi-leptonic BR for $D^+$ decays, to get rid of the uncertainty in the charm quark mass. The function $I$ is the phase space suppression factor and is given by,
 	\begin{equation}
 	I(x)=1-8x+8x^3-x^4+12x^2ln\left(\frac{1}{x}\right).
 	\end{equation}
 	
 	 In the case of $b\to s\gamma$ decay, up-type quarks flow in the loop and the heavy top quark contribution dominates and induces penguin operators already at the electroweak scale. In contrast, all the down 
type quarks are massless compared to the electroweak scale, resulting in no penguin contribution at  this scale within the SM for the case of  $c\to u\gamma$. However, the presence of a heavy down type vector-like quark, will result in a non-vanishing penguin contribution at the electroweak scale. Within the SM, the enhancement of the radiative decay rates in presence of QCD corrections was pointed out in Ref.~\cite{Burdman}. While the enhancement was by 
a factor of two in the case of $b\to s\gamma$, it was expected to be more dramatic in the case of charm radiative decays~\cite{Burdman,greub}. It is hence important to write down the weak effective Hamiltonian with all the 
dimension-6 operators and calculate the corresponding Wilson coefficients within the renormalization-group improved perturbation theory which are discussed below. 

\subsection{The RG evolution and the coefficient $\text{C}_{7_{\text{eff}}}$}\label{Wilson}
The RG evolution of the Wilson coefficients for charm decays in context of the SM
% and a heavy vector-like $q=-1/3$ quark NP model
to the next-to-leading order (NLO) in QCD corrections is performed. The calculation for the complete set of operators relevant for charm decays had previously been done up to NLO in the NDR scheme~\cite{fajfer_wc} and to the next-to-next-to-leading order (NNLO) in the $\overline{\text{MS}}$ scheme~\cite{muller_wc}. In this article, we work in the $\overline{\text{MS}}$ scheme since the anomalous dimension matrices at the leading order (LO) ($\hat{\gamma}^0_{eff}$) and at NLO ($\hat{\gamma}^1_{eff}$) are readily available in Ref.~\cite{muller_wc}. The short distance evolution of the Wilson coefficients has to be divided into two steps.
%Let us consider the short distance evolution of the Wilson coefficients within the SM first. The calculation has to be divided into %two steps. 
The first task is to integrate out the weak gauge bosons at a scale $\mu\sim M_W$. 
This
is done by calculating the $C_i$'s at the scale $\mu\sim M_W$ by matching the effective theory with five active flavors $q = u$, $d$, $s$, $c$, $b$ onto the full theory. As mentioned earlier, no penguin operators are generated at this point, since all 
the down-type quarks ($d$, $s$ and $b$) are to be treated as massless~\cite{greub} and the GIM mechanism is in full effect. The effective hamiltonian for the scale $m_b<\mu<M_W$ is then given by,
\begin{equation}
 \mathcal{H}_{eff}(m_b<\mu<M_W)=\frac{4G_F}{\sqrt{2}}\sum_{q=d,s,b}V_{cq}^*V_{uq}[C_1(\mu)Q_1^q+C_2(\mu)Q_2^q].
\end{equation}
Here,
\begin{eqnarray}
 Q_1^q&=&\left(\bar{u}_{L}\gamma_{\mu}T^{a}q_{L}\right)\left(\bar{q}_{L}\gamma^{\mu}T^{a}c_{L}\right),\;\;\;\;\;\;\;\;
 Q_2^q=\left(\bar{u}_{L}\gamma_{\mu}q_{L}\right)\left(\bar{q}_{L}\gamma^{\mu}c_{L}\right).
\end{eqnarray}
The effective anomalous dimension matrix $\hat{\gamma}_{eff}$ is calculated in the effective theory with five flavours. Using this matrix, the $C_i(M_W)$'s are evolved down to the scale $\mu\sim m_b$, and the $C_i(m_b)$'s are 
obtained.

The next step is to integrate out the $b$ quark as an effective degree of freedom at the scale $\mu\sim m_b$. This is accomplished by matching the effective five flavour theory onto the effective theory for four flavours. This 
generates the penguin operators with the Wilson coefficients depending upon $M_W$ solely through the coefficients $C_{1,2}(m_b)$. The effective hamiltonian at the scale $m_c<\mu<m_b$ is then given by
\begin{equation}
\mathcal{H}_{eff}(m_c<\mu<m_b)=\frac{4G_F}{\sqrt{2}}\sum_{q=d,s}V_{cq}^*V_{uq}[C_1(\mu)Q_1^q+C_2(\mu)Q_2^q+\sum_{i=3}^{10}C_i(\mu)Q_i]
\end{equation}
where
\begin{eqnarray}
 Q_3&=&\bar{u}_{L}\gamma_{\mu}c_{L}\sum_{q=u,d,s,c}\bar{q}\gamma^{\mu}q,\;\;\;\;\;\;\;\;\;
 Q_4=\bar{u}_{L}\gamma_{\mu}T^{a}c_{L}\sum_{q=u,d,s,c}\bar{q}\gamma^{\mu}T^{a}q,\\
 Q_5&=&\bar{u}_{L}\gamma_{\mu}\gamma_{\nu}\gamma_{\rho}c_{L}\sum_{q=u,d,s,c}\bar{q}\gamma^{\mu}\gamma^{\nu}\gamma^{\rho}q,\;\;\;\;\;\;\;\;\;
 Q_6=\bar{u}_{L}\gamma_{\mu}\gamma_{\nu}\gamma_{\rho}T^{a}c_{L}\sum_{q=u,d,s,c}\bar{q}\gamma^{\mu}\gamma^{\nu}\gamma^{\rho}T^{a}q,\\
 Q_7&=&-\frac{g_{em}}{16\pi^2}m_c\bar{u}_{L}\sigma^{\mu\nu}c_{R}F_{\mu\nu},\;\;\;\;\;\;\;\;\;\;
 Q_8=-\frac{g_{s}}{16\pi^2}m_c\bar{u}_{L}\sigma^{\mu\nu}T^{a}c_{R}G^{a}_{\mu\nu}.
\end{eqnarray}
In all of the above, $q_L=P_Lq$ and $P_{R,L}=(1\pm\gamma_5)/2$ are the chirality projection operators. The $T^a$ 
are the generators of $SU(3)$. The $C_i$'s are the Wilson coefficients which contain the complete short distance (perturbative QCD) corrections. For the case of radiative charm decays under consideration here, 
the operators $Q_9$ and $Q_{10}$ are not relevant and therefore not shown in the above list. We will hence consider only the set of Wilson coefficients $C_{1,...,8}$ which are evolved down from the $m_b$ scale to the $m_c$ scale using the $\hat{\gamma}_{eff}$ matrix now evaluated in the effective theory with four flavours to obtain the $C_i(m_c)$'s.

Hence, at each order $(O)$, the vector of the Wilson coefficients $C_i$ at the scale $\mu=m_c$ may be schematically written as
\begin{equation}
 C_i^{(O)}(m_c)=U^{(O)}_{(f=4)}(m_c,m_b)R^{(O)}_{match}U^{(O)}_{(f=5)}(m_b,M_W)C_i^{(O)}(M_W)
\end{equation}
where $f$ is the number of active flavours at the corresponding scale, $R^{(O)}_{match}$ is the matching matrix between the effective five flavour theory above the scale $\mu=m_b$ to the effective four flavour theory below the scale 
$\mu=m_b$, the index $O=$\{LO, NLO\} specifies the order in QCD corrections at which the corresponding quantities are being calculated and the $U$'s are the evolution matrices related to the effective anomalous dimension matrix 
$\hat\gamma_{eff}$ and are discussed in detail below. We use the formalism given in Ref.~\cite{Buchalla_Buras,ref_12_muller_wc} to obtain the evolution matrices for LO and NLO. We also closely follow Ref.~\cite{muller_wc} in the following discussion.
\subsection{The leading order(LO) evolution}
 Let us start with the full $8\times8$ effective anomalous dimension matrix at the leading order ($\hat\gamma_{eff}^0$) which can be assimilated in parts from~\cite{muller_wc,ref_7_muller_wc,ref_10_muller_wc,ref_11_muller_wc}. 
 It is given in eqn.~(\ref{g0}) in appendix~\ref{LONLOanomalous} with the full dependence on the number of active flavours($f$) and charges($q_1, q_2$) of the internal quark and the decaying quark. 

Now, let $V$ be the matrix that diagonalizes $\hat\gamma_{eff}^{0^T}$, so that
\begin{equation}\label{VLO}
 V^{-1}\hat\gamma_{eff}^{0^T}V=\left[\hat\gamma_{{eff}_i}^{(0)^T}\right]_{diag}.
\end{equation}
The LO evolution matrix $U^{(0)}$ for evolving the $C_i$'s down from the scale $\mu_2$ to $\mu_1$ is then given by
\begin{equation}
 U^{(0)}(\mu_1,\mu_2)=V\left[\left(\frac{\alpha_s(\mu_1)}{\alpha_s(\mu_2)}\right)^{-\hat\gamma_{{eff}_i}^{(0)}/2\beta_0}\right]_{diag}V^{-1}
\end{equation}
where $\alpha_s$ is the strong coupling constant. 

A few comments are in order at this point. It was specified previously that the only operators relevant for the case of charm decays within the SM, above the scale $\mu=m_b$ are $Q_1^q$ and $Q_2^q$. Hence, the matrix 
$U^{(0)}(m_b,M_W)$ is essentially a $2\times2$ matrix. The LO values of $C_{1,2}(M_W)$, which are basically the initial conditions are well known and are given by:
\begin{equation}
 C_1(M_W)=0,\;\;\;\;\;\;\;\;C_2(M_W)=1.
\end{equation}
Hence we have, for the scale $m_b<\mu<M_W$
\begin{equation}
 \begin{pmatrix}
  C_1(m_b)\\C_2(m_b)
 \end{pmatrix}=U^{(0)}(m_b,M_W)\begin{pmatrix}
                                C_1(M_W)\\C_2(M_W)
                               \end{pmatrix}.  
\end{equation}
At this point, all the other Wilson coefficients ($C_3$ to $C_8$) are zero. They get their values from the matching at the scale $m_b$. However, the matching matrix $R_{match}=\delta_{ij}$ to LO and hence, for the LO evolution, the 
coefficients $C_3$ to $C_8$ remain vanishing even after the matching procedure. The resulting $8\times1$ column vector $(C_1(m_b),C_2(m_b),0,0,0,0,0,0)$ is then multiplied with the $8\times8$ evolution matrix $U^{(0)}(m_c,m_b)$ to obtain
the values of the $C_i$'s at the charm scale. The renormalization scheme independent Wilson coefficient $C_{7_{eff}}$ relevant for radiative charm decays is then obtained at LO using the relation
\begin{equation}\label{c7eff}
 C_{7_{eff}}=C_7+\sum_{i=1}^{6}y_iC_i
\end{equation}
where $y_i=\frac{2}{3}\{0,0,1,\frac{4}{3},20,\frac{80}{3}\}$~\cite{muller_wc}.

\subsection{The next-to-leading order(NLO) evolution}
The NLO expression for the evolution matrix is given by
\begin{equation}\label{NLO}
 U^{(1)}(\mu_1,\mu_2)=(1+\alpha_s(\mu_1)J^{(1)})U^{(0)}(\mu_1,\mu_2)(1-\alpha_s(\mu_2)J^{(1)})
\end{equation}
where
\begin{equation}
 J^{(1)}=VH^{(1)}V^{-1}.
\end{equation}
$V$ was defined previously in eqn.~(\ref{VLO}) and the matrix $H$ is defined by
\begin{equation}
 H_{ij}^{(1)}=\delta_{ij}\hat\gamma_{{eff}_i}^{(0)}\frac{\beta_1}{2\beta_0^2}-\frac{G_{ij}^{(1)}}{2\beta_0+\hat\gamma_{{eff}_i}^{(0)}-\hat\gamma_{{eff}_j}^{(0)}}.
\end{equation}
with 
\begin{equation}
 G^{(1)}=V^{-1}\hat\gamma_{eff}^{(1)^T}V.
\end{equation}
The expression for the $8\times8$ $\hat\gamma_1$ matrix with the complete effective flavour and charge dependence can again be collected in parts from~\cite{muller_wc,ref_7_muller_wc,ref_10_muller_wc,ref_11_muller_wc}. Due to its
large size, we provide the matrix in two separate $8\times6$ and $8\times2$ blocks in appendix~\ref{LONLOanomalous} (see eqns.~\ref{g11} and~\ref{g12}).  

It is easy to see that one encounters a term of the order of $\alpha_s^2$ on expanding the expression for $U^{(1)}(\mu_1,\mu_2)$ (eqn.~(\ref{NLO})). However, a calculation of the NLO contribution necessarily requires that all 
terms higher than the first order in $\alpha_s$ be discarded and hence, special care should be taken in using eqn.(\ref{NLO}) for the NLO evolution.
 
Similar to the case of the LO evolution, the only relevant coefficients above the $m_b$ scale are $C_1(M_W)$ and $C_2(M_W)$, calculated up to the NLO order this time. The expressions can be found in~\cite{Misiak} and in the 
$\overline{\text{MS}}$ scheme are given by
\begin{equation}
 C_1(M_W)=\frac{15\alpha_s(M_W)}{4\pi},\;\;\;\;\;\;\;\;\;\;\;\;\;\;\;\;C_2(M_W)=1.
\end{equation}
The coefficients $C_i(i=3,...,8)$ however are non-vanishing after the matching procedure at NLO, since the matching matrix $R_{match}$ is now defined by
\begin{equation}
 R_{{match}_{ij}}=\delta_{ij}+\frac{\alpha_s(m_b)}{4\pi}R_{ij}^{(1)}.
\end{equation}
The non-zero elements of the matrix $R^{(1)}$ for charm decays being~\cite{muller_wc}
 \begin{eqnarray}
 R_{41}^{(1)}&=&-R_{42}^{(1)}/6=1/9,\nonumber\\
 R_{71}^{(1)}&=&-R_{72}^{(1)}/6=8/81,\nonumber\\
 R_{81}^{(1)}&=&-R_{82}^{(1)}/6=-1/54.
 \end{eqnarray}

The full set of NLO coefficients $(C_1,...,C_8)$ for the case of charm decays in the SM is then given by
\begin{equation}
  C(m_c)=U^{(1)}(m_c,m_b)R_{match}C(m_b).  
\end{equation}
where $C(m_b)$ is an $8\times1$ column vector whose first two elements are $C_1(m_b)$ and $C_2(m_b)$ and the rest are zero. Once the values at the charm scale are obtained, the corresponding value for $C_{7_{eff}}$ can be obtained from eqn.~(\ref{c7eff}).

\section{New Physics models}
\label{III}
\subsection{Down type isosinglet vector-like quark}\label{VLQ}
The SM contains three generation of quarks, however, the number of generations is not predicted by the theory. A simple extension of the SM would be to have a chiral fourth generation of quarks and leptons. Presence of a fourth 
generation would have a significant effect on the Higgs sector of the SM and is now ruled out by the Higgs production and decay processes data at the LHC. However, the so called vector-like quarks, which do not receive their 
masses from Yukawa couplings to a Higgs doublet, are consistent with the present Higgs data. They are distinguished from the SM quarks by their vector coupling to gauge bosons, i.e., both the left handed, $\Psi_L$ and right 
handed, $\Psi_R$ chiralities of these fermions transform the same way under the SM gauge groups $SU(3)_c\times SU(2)_L\times U(1)_Y$. These exotic fermions occur for example in the grand unified theory based on 
$E_6$~\cite{HewettRizzo}. In general these fermions could either be singlets or doublets or triplets under $SU(2)_L$. Here we consider the case of a down type isosinglet quark.
%Constraints on mass and mixings 
In the SM, the quark mixing matrix is a $3\times 3$ unitary matrix which is specified in terms of three angles ($\theta_{12},\theta_{13},\theta_{23}$) and a $CP$-violating phase, $\delta_{13}$. A $4\times 4$ unitary quark mixing 
matrix is parametrized in terms of 3 additional angles ($\theta_{14},\theta_{24},\theta_{34}$) and two more $CP$ violating phases, $\delta_{14},\delta_{24}$. In Ref.~\cite{uma}, a chi-squared fit to many flavour observables 
was performed to obtain the preferred central values, along with the errors of all the elements of the measurable $3\times 4$ quark mixing matrix. To evaluate the SD contribution of the radiative decay rate in the presence of the vector-like isosinglet quark, the central values of the mass and mixing angles are obtained from the results of the fit in Ref.~\cite{uma} are used.
%There are several models with the Higgs doublet being standard model like, with the vector-like quarks being iso-singlets, doublets or triplets.     
   
\subsection*{Modified Wilson coefficients in presence of a vector-like quark}\label{vector_like}
Having discussed the evolution of the Wilson coefficients for the SM in full detail we will now simply specify how the contribution of the vector-like quark model modifies the SM coefficients.

%The NP model that we use is a heavy vector-like quark model with a down-type isosinglet ($q=-1/3$) heavy quark, which we denote by $b^\prime$.
The down-type vector-like quark induces a $Z$-mediated FCNC in the down-type quark sector. In Ref.~\cite{Alok_Uma} it was pointed out for the case of singlet up type vector like quark, that only the Wilson coefficients
are modified. Similarly for the
$c\rightarrow u$ transitions that are of interest to us, no new set of operators are introduced and hence the anomalous dimension matrices along with the coefficients $C_1(M_W)$ and $C_2(M_W)$, remain exactly the same as that in 
the SM, up to NLO.\footnote{However, at the NNLO order, one encounters terms dependent on $\frac{m_t^2}{M_W^2}$ which arise as a result of integrating out the top quark as a heavier degree of freedom at the electroweak scale. 
Since the 
	$b^\prime$ is also heavier than the $W$ boson, one needs to integrate it out too at this scale. Hence, at the NNLO level, the expressions for $C_1(M_W)$ and $C_2(M_W)$ change for this model as 
	compared to SM.} 

The fundamental difference in this model is that at the electroweak scale, the coefficients $C_{7,8}$ will not be zero. While the down-type quarks running in the penguin loop in the SM\footnote{The relevant diagrams in the Feynman gauge can be found in~\cite{Kim}.} can be treated as massless and hence do not contribute, the vector-like $b^{\prime}$ quark, which couples with all the up-type SM quarks being heavier than $M_W$ will generate a value for the coefficients $C_7$ and $C_8$ at the electroweak scale itself. The values are
\begin{eqnarray}
	C_7&=&\frac{1}{2}\left(G_1\left(\frac{m_{b^\prime}^2}{M_W^2}\right)-\frac{1}{3}G_2\left(\frac{m_{b^\prime}^2}{M_W^2}\right)\right)\\
	C_8&=&\frac{1}{2}G_2\left(\frac{m_{b^\prime}^2}{M_W^2}\right)
\end{eqnarray}
where the functions $G_p(r)$ defined in~\cite{Kim} are given in appendix~\ref{G}.

We have calculated these coefficients in this model for two benchmark values for the mass of the $b^{\prime}$ quark in accordance with~\cite{uma}. Our results are displayed in Table~\ref{tab1}. Our values for the coefficients in the SM match exactly with Ref.~\cite{muller_wc} if we use their values for the parameters $m_t$, $m_b$, $M_W$ and $\mu$. We find there is more than an order
enhancement in the values of the coefficients $C_{7_{eff}}$ and $C_{8_{eff}}$ at the NLO level in the case of this vector-like quark model compared to the SM. However, we should mention here that our NLO results for the NP model are not exact in the sense 
that we have not calculated the expressions for these coefficients at the NLO level at the $W$ scale. The LO results are exact.
\begin{table*}[ht]
	\centering
	\caption{The values of the Wilson coefficients at the charm scale in SM and a heavy vector-like quark(VLQ) model with the benchmark values of $800$ GeV and $1200$ GeV for the heavy-quark mass. We take the mass of 
	the charm quark $m_c=1.275$ GeV, the $\overline{\text{MS}}$ mass of the bottom quark $m_b=4.18$ and the mass of the $W$ boson $M_W=80.385$. The four-loop expression for the strong constant $\alpha_s$ has been used.}
	\begin{tabular}{|c|c|c|c|c|c|c|}
		\hline
		% \multirow{2}{*}{Object} & \multicolumn{1}{c|}{\multirow{2}{*}{\textbf{Action}}} &  \multicolumn{6}{c|}{\textbf{RMS Errors}} & \textbf{Leave/Off} & \multirow{2}{*}{\textbf{Total}} \\
		\multirow{2}{*}{\textbf{Coefficients}}& \multicolumn{3}{c|}{\textbf{LO}}& \multicolumn{3}{c|}{\textbf{NLO}} \\
		\cline{2-7}
		&\textbf{SM} & \textbf{VLQ}   & \textbf{VLQ}    &\textbf{SM} & \textbf{VLQ}   & \textbf{VLQ}   \\
		&            & $m_{b^\prime}=800$ GeV & $m_{b^\prime}=1200$ GeV &            & $m_{b^\prime}=800$ GeV & $m_{b^\prime}=1200$ GeV \\   \hline
		$C_1$         & -1.0769 & -1.0769 & -1.0769 & -0.7434 & -0.7434 & -0.7434 \\ 
		\hline
		$C_2$         &  1.1005 &  1.1005 &  1.1005 &  1.0503 &  1.0503 & 1.0503  \\ 
		\hline
		$C_3$         & -0.0043 & -0.0043 & -0.0043 & -0.0060 & -0.0060 & -0.0060 \\ 
		\hline
		$C_4$         & -0.0665 & -0.0665 & -0.0665 & -0.1015 & -0.1015 & -0.1015 \\ 
		\hline
		$C_5$         &  0.0004 &  0.0004 &  0.0004 &  0.0003 &  0.0003 &  0.0003 \\ 
		\hline
		$C_6$         &  0.0008 &  0.0008 &  0.0008 &  0.0009 &  0.0009 &  0.0009 \\ 
		\hline
		$C_7$         &  0.0837 &  0.3324 &  0.3276 &  0.6095 &  0.2820 &  0.2778 \\ 
		\hline
		$C_8$         & -0.0582 & -0.2259 & -0.2253 & -0.0690 & -0.2197 & -0.2192 \\ 
		\hline
		$\mid C_{7_{eff}}\mid$ &  0.0424 &  0.2911 &  0.2863 & 0.0119  &  0.2159  & 0.2117\\ 
		\hline
	\end{tabular}
	\label{tab1}
\end{table*}  
From the values in Table~\ref{tab1} it is evident that the dimension six operators $O_{1,...,6}$ do not mix with the dimension five operators $O_{7,8}$ (a fact that is well known and clear from the form of the anomalous dimension 
matrices). 
\subsection{Left-right symmetric model}\label{LRSM}
The minimal Left Right symmetric model is based on the gauge group $SU(3)_{c}\times SU(2)_{L}\times SU(2)_{R}\times U(1)_{B-L}$~\cite{LR_model_1,LR_model_2,LR_model_3} with the fermions represented as doublet representations of $SU(2)_{L}$ and $SU(2)_{R}$. The electric charge $Q$ and the third components of the weak isospin $I_{3L}$ and $I_{3R}$ are related as $Q=I_{3L}+I_{3R}+\frac{B-L}{2}$. To ensure perturbative interactions between right-handed gauge boson and fermions, $\zeta_{g}=\frac{g_{R}}{g_{L}}$ (where the $g_R$ and $g_L$ are the right and left handed couplings respectively) should not be large. As in the low energy weak interaction
L-R symmetry is broken, in general $g_{L}\neq g_{R}$. Direct search results impose the the bound $\zeta_{g}M_{W_{2}}>2.5$ TeV~\cite{G.Aad,Chatrchyan}. In order
to generate active neutrino mass through see-saw mechanism, $v_{R}$ should be in the TeV range. All these constraints result in the range for $\zeta_{g}$ being $[0,2]$. The charged gauge boson $W_{L}$ and $W_{R}$ are mixture of the mass eigenstates $W_{1}$ and $W_{2}$, with a mixing angle $\zeta$ restricted to lie in the range $[0,10^{-3}]$~\cite{Cho,Kou}.

The effective lagrangian given in eqn.~(\ref{SM_Lag}) (for SM) may now be written for the case of LRSM as,
\begin{equation}
 \mathcal{L}_{eff}=-\frac{eG_{F}}{4\sqrt{2}\pi^{2}}\big[\mathcal{A}\bar{u}\sigma^{\mu\nu}RcF_{\mu\nu}+\mathcal{B}\bar{u}\sigma^{\mu\nu}LcF_{\mu\nu}\big]
\end{equation}
where $\mathcal{A}$ and $\mathcal{B}$ are the bare SD contributions to $c_L$ and $c_R$ respectively and are given by~\cite{to_come}
\begin{align}
\mathcal{A}&=\sum_{\ell}\bigg\{Q_{1}\big(M\text{cos}^{2}\zeta\lambda_{\ell}^{LL}G_{1}^{LL}+m\zeta_{g}^{2}\text{sin}^{2}\zeta\lambda_{\ell}^{RR}G_{1}^{RR}+
m_{\ell}\zeta_{g}\text{sin}\zeta\text{cos}\zeta e^{i\phi}\lambda_{\ell}^{LR}G_{1}^{LR}\nonumber\\
&+m_{\ell}\zeta_{g}\text{sin}\zeta\text{cos}\zeta e^{-i\phi}\lambda_{\ell}^{RL}G_{1}^{RL}\big)
+Q_{2}\big(M\text{cos}^{2}\zeta\lambda_{\ell}^{LL}G_{2}^{LL}
+m\zeta_{g}^{2}\text{sin}^{2}\zeta\lambda_{\ell}^{RR}G_{2}^{RR}\nonumber\\
&+m_{\ell}\zeta_{g}\text{sin}\zeta\text{cos}\zeta e^{i\phi}\lambda_{\ell}^{LR}G_{2}^{LR}
+m_{\ell}\zeta_{g}\text{sin}\zeta\text{cos}\zeta e^{-i\phi}\lambda_{\ell}^{RL}G_{2}^{RL}\big)\bigg\}\label{AB_amps_1}\\
\mathcal{B}&=\sum_{\ell}\bigg\{Q_{1}\big(m\text{cos}^{2}\zeta\lambda_{\ell}^{LL}H_{1}^{LL}+M\zeta_{g}^{2}\text{sin}^{2}\zeta\lambda_{\ell}^{RR}H_{1}^{RR}+
m_{\ell}\zeta_{g}\text{sin}\zeta\text{cos}\zeta e^{i\phi}\lambda_{\ell}^{LR}H_{1}^{LR}\nonumber\\
&+m_{\ell}\zeta_{g}\text{sin}\zeta\text{cos}\zeta e^{-i\phi}\lambda_{\ell}^{RL}H_{1}^{RL}\big)
+Q_{2}\big(m\text{cos}^{2}\zeta\lambda_{\ell}^{LL}H_{2}^{LL}
+M\zeta_{g}^{2}\text{sin}^{2}\zeta\lambda_{\ell}^{RR}H_{2}^{RR}\nonumber\\
&+m_{\ell}\zeta_{g}\text{sin}\zeta\text{cos}\zeta e^{i\phi}\lambda_{\ell}^{LR}H_{2}^{LR}
+m_{\ell}\zeta_{g}\text{sin}\zeta\text{cos}\zeta e^{-i\phi}\lambda_{\ell}^{RL}H_{2}^{RL}\big)\bigg\}\label{AB_amps_2}.
\end{align}
For the case of $c\rightarrow u\gamma$ decays, $Q_{1}=1, Q_{2}=-1/3, M=m_c$ and $m=m_u$. $\lambda_l$'s are the CKM factors, $\lambda_{\ell}^{LL}=V_{c\ell}^{*L}V_{\ell u}^{L}$, $\lambda_{\ell}^{RR}=V_{c\ell}^{*R}V_{\ell u}^{R}$,
$\lambda_{\ell}^{LR}=V_{c\ell}^{*L}V_{\ell u}^{R}$, $\lambda_{\ell}^{RL}=V_{c\ell}^{*R}V_{\ell u}^{L}$.
The mass of the down-type quarks running in the penguin loop is represented by $m_l$. The functions $G_{p}^{ij}$ and $H_{p}^{ij}$ 
for $p=1,2$ and $i=j=L$ are given in Ref.~\cite{Kim}. $G_p$ is also included in appendix~\ref{G}. The $i=j=R$, $i=L$, $j=R$ and the $i=R$, $j=L$ counterparts relevant for the LRSM are explained in detail in Ref.~\cite{to_come}. We calculate the SD contributions $\mathcal{A}$ and $\mathcal{B}$ only at the bare level. For the $c\rightarrow u\gamma$ decays in the LRSM model, the operator basis with and without the heavy vector-like quark now consists of 20 operators. They are the 8 operators described in sec.~\ref{Wilson} which contribute to $\mathcal{A}$ along with the following two operators, 
\begin{eqnarray} 
 Q_9^q&=&\left(\bar{u}_{L}\gamma_{\mu}T^{a}q_{L}\right)\left(\bar{q}_{R}\gamma^{\mu}T^{a}c_{R}\right),\;\;\;\;\;\;\;\;
 Q_{10}^q=\left(\bar{u}_{L}\gamma_{\mu}q_{L}\right)\left(\bar{q}_{R}\gamma^{\mu}c_{R}\right),
\end{eqnarray}
which are the left-right analogues of $Q_1^q$ and $Q_2^q$.
10 more operators with the chiralities of these operators flipped, contribute to $\mathcal{B}$. Since the strong interactions preserve chirality, these two sets of operators with different chiralities do not mix with each other and the RG group mixing of the two sets are the same. However, the additional operators require an additional $\gamma_{4\times4}$ which although present in the literature for radiative $b$ decays~\cite{Cho}, is not available for the case of the radiative charm decays. Hence incorporating the QCD corrections for the LRSM case, is beyond the scope of this work.
\section{Results and discussions}
\label{IV}
\subsection{Branching ratios in the SM and for the NP models}\label{IVA}
 The inclusion of QCD corrections result in an enhancement of the coefficient $A^{SM}$ (defined in eqn.~\ref{Asm}) from $\mathcal{O}(10^{-7})$ at the bare (QCD uncorrected) level to $\mathcal{O}(10^{-6})$ at the LO and $\mathcal{O}(10^{-3})$ at the NLO level. At the LO, the contributions from the intermediate $d$ and $s$ quarks differ only in the CKM factors $V_{cd}^*V_{ud}$ and $V_{cs}^*V_{us}$. Their sum, using unitarity is $-V_{cb}^*V_{ub}$, leading to a large suppression in the amplitude. At the NLO, the functional dependence of the amplitudes on the $s$ and $d$ quark masses becomes substantial and hence the net amplitude is no longer just the sum of the CKM factors. In fact, since $V_{cs}^*V_{us}=-V_{cd}^*V_{ud}$, this results in $A^{SM}\propto V_{cs}^*V_{us}[f(\frac{m_s}{mc})^2-f(\frac{m_d}{mc})^2]$, where the function $f$~\cite{greub} is given by:
\begin{align}
f(x)= &-\frac{1}{243}((3672-288\pi^{2}-1296\zeta_{3}+(1944-324\pi^{2})\text{ln}\,x+108\text{ln}^{2}\,x+36\text{ln}^{3}\,x)x+576\pi^{2}x^{\frac{3}{2}}\nonumber\\
 &+(324-576\pi^{2}+(1728-216\pi^{2})\text{ln}\,x+324\text{ln}^{2}\,x+36\text{ln}^{3}\,x)x^{2}+(1296-12\pi^{2}+1776\text{ln}\,x\nonumber\\
 &-2052\text{ln}^{2}\,x)x^{3}) -\frac{4\pi i}{81}((144-6\pi^{2}+18\text{ln}\,x+18\text{ln}^{2}\,x)x+(-54-6\pi^{2}+108\text{ln}\,x+18\text{ln}^{2}\,x)\nonumber\\
 &x^{2}+(116-96\text{ln}\,x)x^{3}).
\end{align}
Hence, the coefficient $A^{SM}$ at LO and NLO is given by,
\begin{align}
 &A^{SM}_{LO}=-V_{cb}^{*}V_{ub}C_{7_{eff}}^{LO}\;\;\;\;\;\;\;\;\;\;
 &A^{SM}_{NLO}=V_{cs}^{*}V_{us}C_{7_{eff}}^{NLO}.
\end{align} 
 Note that $\mid C_{7_{eff}}\mid$ itself is not enhanced at NLO compared to LO within the SM as is evident from the values in Table~\ref{tab1}, rather the different CKM coefficients appearing in $A^{SM}_{LO}$ and $A^{SM}_{NLO}$ result in the enhancement of the coefficient $A^{SM}$ at the NLO level. 
 
 Since the vector-like quark $b^\prime$ generates a non-vanishing value for the coefficients $C_7$ and $C_8$ at the electroweak scale itself, its presence results in an increased magnitude of $C_{7_{eff}}$ as can be seen in
 Table~\ref{tab1}. This results in the BR enhancement by 2 orders of magnitude in the vector-like quark model at NLO compared to that in the SM. The values for $\mid A\mid$ and the corresponding BR's for the QCD uncorrected, LO and NLO corrected contributions for SM and the vector-like quark model (with $m_b^\prime=800, 1200$ GeV) are given in table~\ref{tab2}.  
\begin{table*}
	\centering
	\caption{The values for $\mid A\mid$ and the inclusive $c\rightarrow u\gamma$ BR in the SM and vector-like quark(VLQ) model. For the vector-like quark model, the values have been calculated for the benchmark values $m_b^\prime=800$ GeV and $1200$ GeV.}
	\begin{tabular}{|c|c|c|c|c|c|c|}
		\hline
		% \multirow{2}{*}{Object} & \multicolumn{1}{c|}{\multirow{2}{*}{\textbf{Action}}} &  \multicolumn{6}{c|}{\textbf{RMS Errors}} & \textbf{Leave/Off} & \multirow{2}{*}{\textbf{Total}} \\
		\multirow{2}{*}{QCD corrections}& \multicolumn{3}{c|}{$\mid A\mid$}& \multicolumn{3}{c|}{BR($c\rightarrow u\gamma$)} \\
		\cline{2-7}
		&\textbf{SM} & \textbf{VLQ}   & \textbf{VLQ}    &\textbf{SM} & \textbf{VLQ}   & \textbf{VLQ}   \\
		&            & $m_{b^\prime}=800$ GeV & $m_{b^\prime}=1200$ GeV &            & $m_{b^\prime}=800$ GeV & $m_{b^\prime}=1200$ GeV \\   \hline
		Bare         &$2.73\times10^{-7}$&$2.49\times10^{-5}$&$2.35\times10^{-5}$&$2.04\times10^{-17}$&$1.70\times{10^{-13}}$&$1.51\times10^{-13}$ \\ 
		\hline
		LO         &$5.89\times10^{-6}$&$4.32\times10^{-5}$&$4.25\times10^{-5}$&$9.48\times10^{-15}$&$5.11\times{10^{-13}}$&$4.94\times10^{-13}$  \\ 
		\hline
		NLO         &$2.61\times10^{-3}$&$4.46\times10^{-2}$&$4.37\times10^{-2}$&$1.86\times10^{-9}$&$5.46\times{10^{-7}}$&$5.23\times10^{-7}$ \\ 
		\hline
	\end{tabular}
	\label{tab2}
\end{table*}
\begin{table*}
	\centering
\caption{Branching ratios for the LRSM model without and with contribution from heavy vector-like quark(VLQ). The Branching ratio is expressed as a function of $\zeta$, $\zeta_g$ and $\theta_{12}$ (for LRSM) and of $\zeta$, $\zeta_g$, $\theta_{12}$, $\theta_{14}$, $\theta_{24}$ and $\theta_{34}$ (for LRSM+VLQ). The corresponding parameters are varied to determine the maximum and minimum values.}
\begin{tabular}{|c|c|c|}\hline
~Model~&~~&~BR\\
\hline\hline 
LRSM                &Max &$1.96\times10^{-11}$\\
                    &Min &$0.67\times10^{-15}$\\\hline
LRSM+VLQ ($800$ GeV) &Max &$4.65\times10^{-8}$\\\
                    &Min &$1.69\times10^{-13}$\\\hline
LRSM+VLQ ($1200$ GeV)&Max &$0.96\times10^{-7}$\\\
                    &Min &$1.42\times10^{-13}$\\\hline
\hline
\end{tabular}
\label{lrsmbr}
\end{table*}

Table~\ref{lrsmbr} shows the bare level BR's for the LRSM as well as LRSM with a heavy vector-like quark model. Comparing with the bare level BR's from Table~\ref{tab2}, it is evident that for LRSM alone an enhancement of $\mathcal{O}(10^2)$ to $\mathcal{O}(10^6)$ is feasible, depending on the values of the parameters of LRSM, compared to the SM. LRSM along with vector-like quark can enhance the BR by even upto $\mathcal{O}(10^{10})$.

For the SM (vector-like quark model), the enhancement of the BR from the bare level to that with QCD corrections at NLO level is $\mathcal{O}(10^8)$ ($\mathcal{O}(10^6)$). For the case of LRSM with vector-like quark, QCD corrections are expected to lead to similar large enhancement. Even if the enhancement from these corrections is considerably less ($\sim\mathcal{O}(10^{4})$), the QCD corrected SD contribution from the LRSM with vector-like quark could result in BR's much larger than that from the LD effects. This enhancement could possibly point towards the presence of such a NP.

%orders of magnitude enhancements are possible within the allowed parameter ranges for these models even at the bare level, as compared to the SM or even the heavy vector-like quark model.

\subsection{Photon polarization as a probe for new physics}
\label{IVB}
Within the SM, in the penguin diagram responsible for the $c\rightarrow u\gamma$ process only left-handed components of the external fermions couple to the $W$ boson. A helicity flip on the $c$ quark leg, proportional to $m_c$, contributes to the amplitude for the emission of left polarized photons, while, that on the $u$ quark leg, proportional to $m_u$, results in right polarized photons.

In the LRSM since the physical $W_1$ boson couples to both left and right handed quarks, a helicity flip is also possible on the internal ($d$, $s$, $b$) quark lines and will result in additional left handed photons with an amplitude involving a new coefficient function and proportional to $m_b\zeta$ and similarly there will be additional right-handed photons proportional to $m_b\zeta$. In the presence of a vector-like quark, each of these contributions will be proportional to $m_{b^\prime}\zeta$.

% Recently LHCb had reported the direct observation of the photon polarization in the $b\to s\gamma$ transition~\cite{LHCb}. Information about the photon polarization was obtained using the angular distribution of the photon 
% direction with respect to the normal to the plane defined by the momenta of the three final state hadrons in their center-of-mass frame. If one considers the exclusive mode, $D^0\to \omega\gamma$, where $\omega$ decays to three 
% pions, a similar up-down asymmetry between the number of events with photon on either side of the plane can be used to determine the photon polarization.     
In analogy to Ref.~\cite{Gronau}, we define the photon polarization for the inclusive process $c\rightarrow u\gamma$ as
\begin{equation}
 \lambda_{\gamma}=\frac{|c_R|^{2}-|c_L|^{2}}{|c_R|^{2}+|c_L|^{2}}~,
\end{equation}
where $c_R$, $c_L$ denote the amplitudes for the right and left polarized photons in the process.

For the SM,
since the SD contributions to $c_{R}$ are negligible ($\mathcal{O}(m_u)$ suppressed), if one only includes the
SD contributions to estimate $\lambda_{\gamma}$, its value would be $-1$. However, the exclusive decay modes corresponding to the $c\rightarrow u\gamma$ process are dominated by LD contributions. To account for these %the LD contributions 
we add the values of the pole type and VMD amplitudes of all the exclusive processes (given in appendix~\ref{LDBR}). %account for the maximum possible LD contribution. 
Due to uncertainty in the sign of the VMD contributions, the long distance amplitudes can lie in the range ($2.08\times 10^{-9}-8.78\times 10^{-7}$) GeV$^{-1}$. %dominant 
Since the LD amplitude does not have any preferred polarization, it contributes equally to both $c_R$ and $c_L$. This results in an almost vanishing value
of $\lambda_{\gamma}\sim$($\mathcal{O}(10^{-8}-10^{-5})$) within the SM. This is in contrast to the $b\rightarrow s\gamma$ case~\cite{Gronau}, where the LD contributions are less significant,
and hence the $\lambda_\gamma$ value is $-1$ in SM.
Without LR symmetry, an isosinglet vector-like quark can only couple to $W_L$ and hence its addition will only enhance the left handed polarized amplitude. For this case we find that in presence of LD contribution,
$\lambda_\gamma$ lies in the range $-6.1\times 10^{-6}$ to $-2.6\times 10^{-3}$. The bare SD contributions to the $\mid c_R\mid$ and $\mid c_L\mid$ amplitudes within the LRSM are given by eqns.~(\ref{AB_amps_1}) and~(\ref{AB_amps_2}). Here again the LD contribution is appropriately added to $\mid c_R\mid$ and $\mid c_L\mid$. 

Since the minimal LRSM models with an exact symmetry between the left and right handed sectors are becoming harder to realize, we use a right handed mixing matrix which is distinct from the left handed CKM matrix. To decrease the number of parameters, we take the right-handed CKM matrix to be,
\begin{equation}
 \begin{pmatrix}
  \cos\theta_{12}&\sin\theta_{12}&0\\
 -\sin\theta_{12}&\cos\theta_{12}&0\\
  0&0&1 
 \end{pmatrix}.
\end{equation}
This parametrization of the right handed CKM matrix is inspired by Ref.~\cite{Langacker, Kou}. The CP violating phases have been taken to be zero and $\theta_{13}=\theta_{23}=0$,
where $\theta_{ij}$ is the mixing angle between the $i^{th}$ and $j^{th}$ generations. The photon polarization can be expressed as a function of $\zeta$, $\zeta_g$ and $\theta_{12}$.
We vary the parameters $\zeta_g$ and $\zeta$ within their allowed ranges ($0\leq\zeta_g\leq2$ and $0\leq\zeta\leq10^{-3}$) and look for the $\theta_{12}$ values for the maximum deviation of
the polarization from its SM value of $\approx 0$. For the case of the LRSM with a heavy vector-like quark $b^\prime$, there are three additional parameters ($\theta_{14}, \theta_{24}, \theta_{34}$),
which are also chosen to get the maximum deviation of $\lambda_\gamma$ from its SM value. % of $0$.

\begin{figure}[ht]
%\centering
\subfloat[][]{\includegraphics[width=0.4\textwidth]{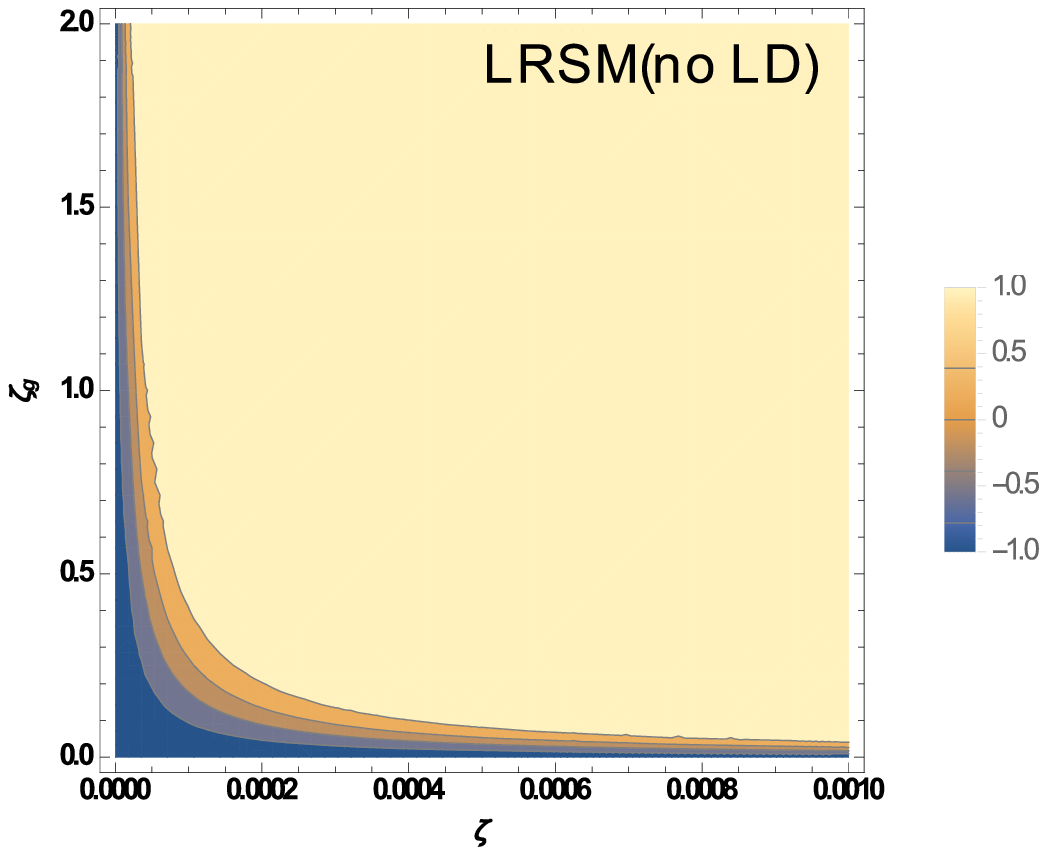}}\hfill
\subfloat[][]{\includegraphics[width=0.41\textwidth]{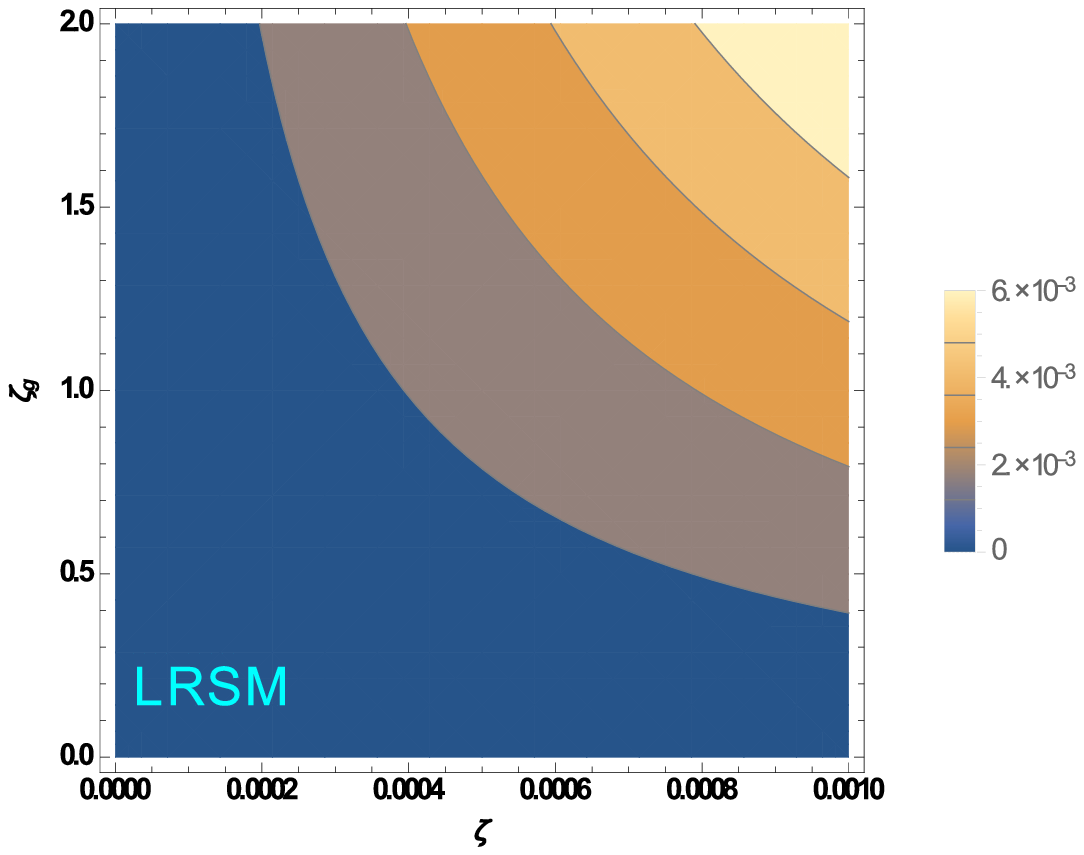}}\\
\caption{\small{Contour plots showing the variation of the polarization $\lambda_\gamma$ as a function of $\zeta$ and $\zeta_g$ for LRSM with no LD contribution on left
and LRSM with LD amplitude of $2\times10^{-9}$ GeV$^{-1}$ on right. For both the cases, the right-handed CKM elements are set for maximum deviation of the polarization function from its SM value.
The bar-legends for the different contours of $\lambda_\gamma$ are displayed along with the respective figures. Here $0<\zeta<10^{-3}$ and $0<\zeta_g<2$}.}
\label{polarization_fig}
\end{figure}

The contour plots for the variation of $\lambda_\gamma$ for LRSM with no LD contribution and LRSM with LD amplitude of $2\times 10^{-9}$ GeV$^{-1}$ are shown in Fig.~\ref{polarization_fig}.
% \begin{figure}
% \begin{subfigure}{0.43\textwidth}
% \includegraphics[width=1\linewidth]{maxdevthetanoqcdLRpart2.jpg}
% \caption{LRSM}
% \end{subfigure}
% \hspace*{\fill}  % spread out the first and second subfigures 
% \begin{subfigure}{0.43\textwidth}
% \includegraphics[width=1\linewidth]{maxdevthetanoqcdLRnoLDpart2.jpg}%
% \caption{LRSM with no LD}
% \end{subfigure}
% 
% \bigskip  % create some vertical separation between the two rows of subfigures
% 
% \begin{subfigure}{0.43\textwidth}
% \includegraphics[width=1\linewidth] {maxdevthetanoqcdNP1part2.jpg}
% \caption{LRSM+vector-like quark $m_{b^\prime}=800$ GeV)
% \end{subfigure}
% \hspace*{\fill} % spread out the third and fourth subfigures 
% \begin{subfigure}{0.43\textwidth}
% \includegraphics[width=1\linewidth] {maxdevthetanoqcdNP2part2.jpg}%
% \caption{LRSM+vector-like quark $m_{b^\prime}=1200$ GeV)}
% \end{subfigure}
% 
% \caption{Contour plots showing the variation of the polarization $\lambda_\gamma$ as a function of $\zeta$ and $\zeta_g$ for (a) LRSM, (b) LRSM with no LD contribution, (c) LRSM with vector-like quark where $m_b^\prime=800$ GeV and (d) LRSM with vector-like quark where $m_b^\prime=1200$ Gev with the right-handed CKM elements set for maximum deviation of the polarization function from $-1$. (e) Shows the barlegend for the different contours of $\lambda_\gamma$. Here $0<\zeta<10^{-3}$ and $0<\zeta_g<2$.} \label{polarization_fig}
\begin{figure}[ht]
%\centering
\subfloat[][]{\includegraphics[width=0.36\textwidth]{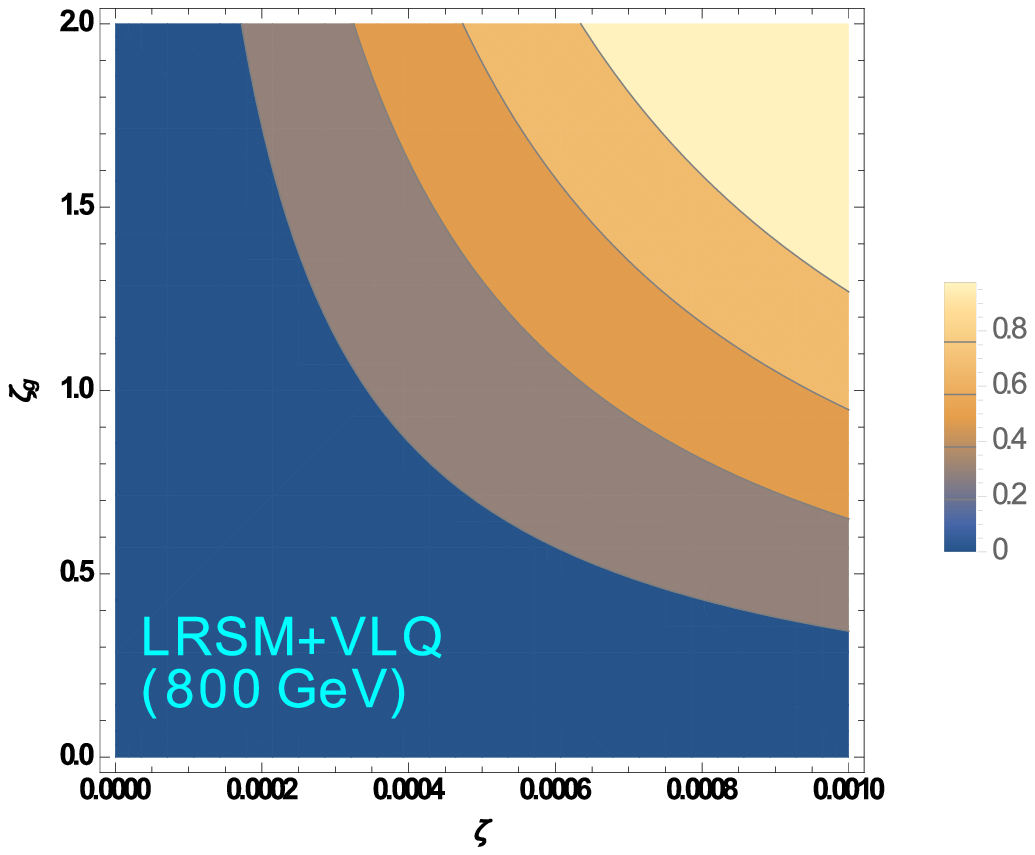}}
\subfloat[][]{\includegraphics[width=0.36\textwidth]{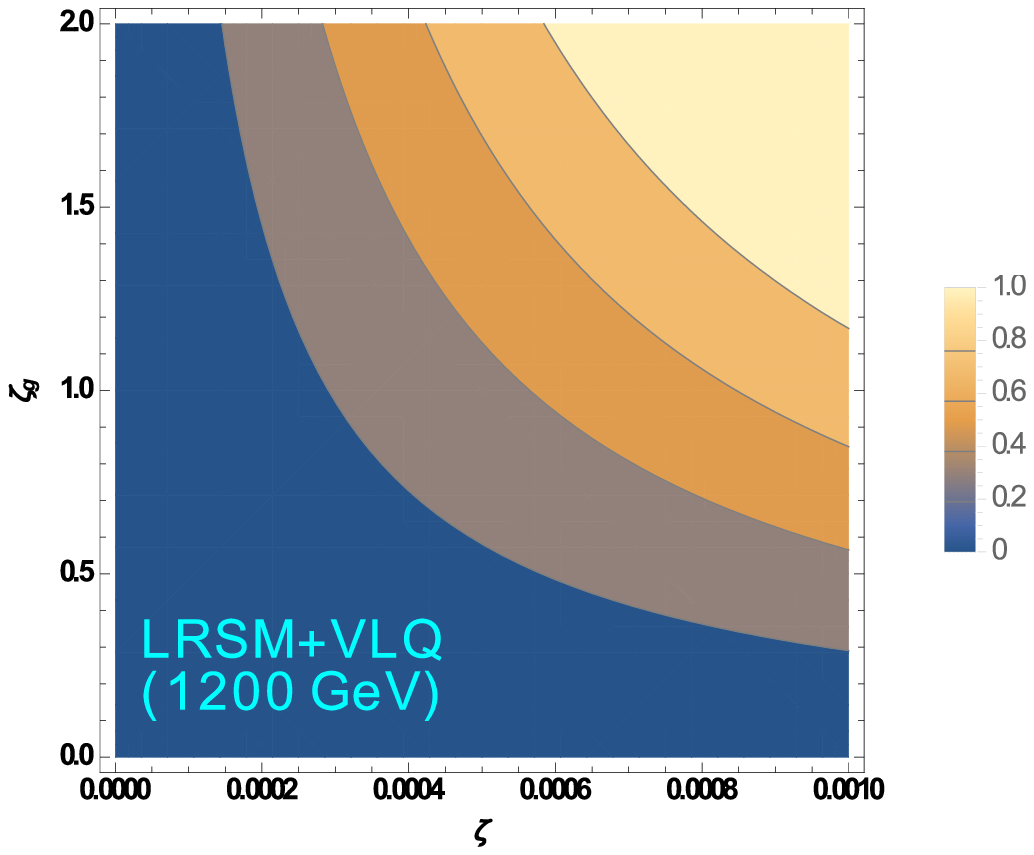}}\\
\subfloat[][]{\includegraphics[width=0.36\textwidth]{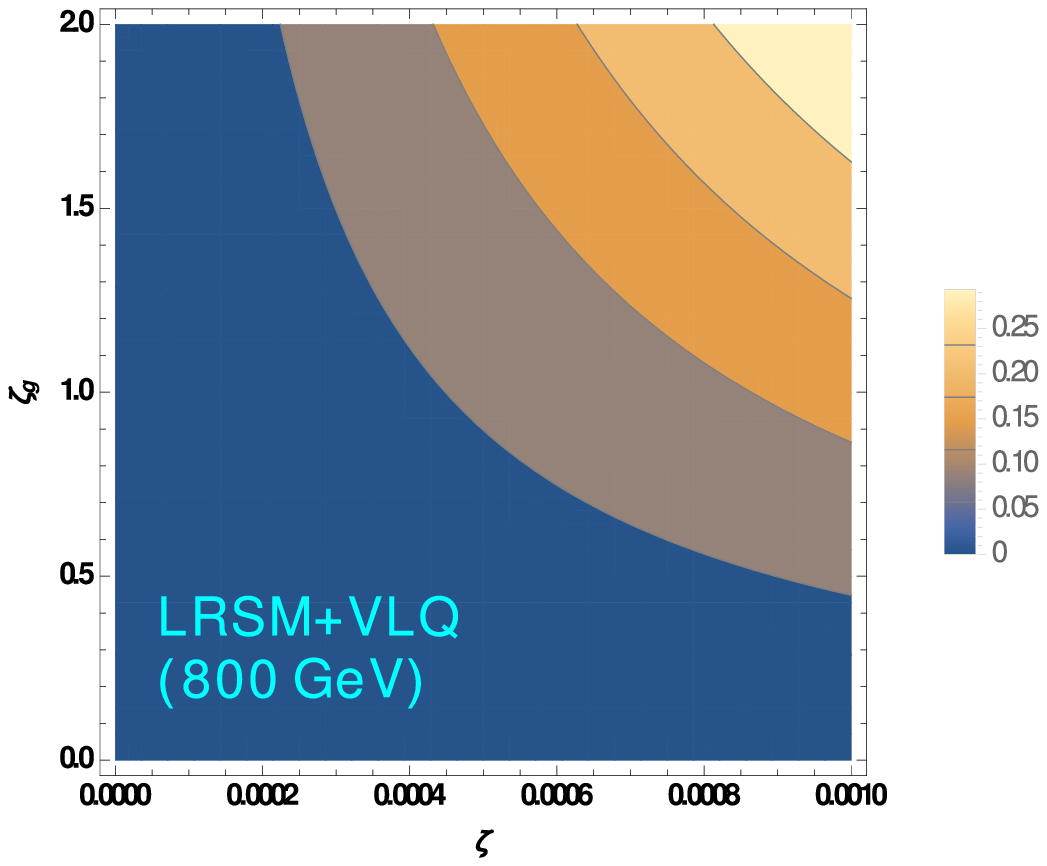}}
\subfloat[][]{\includegraphics[width=0.36\textwidth]{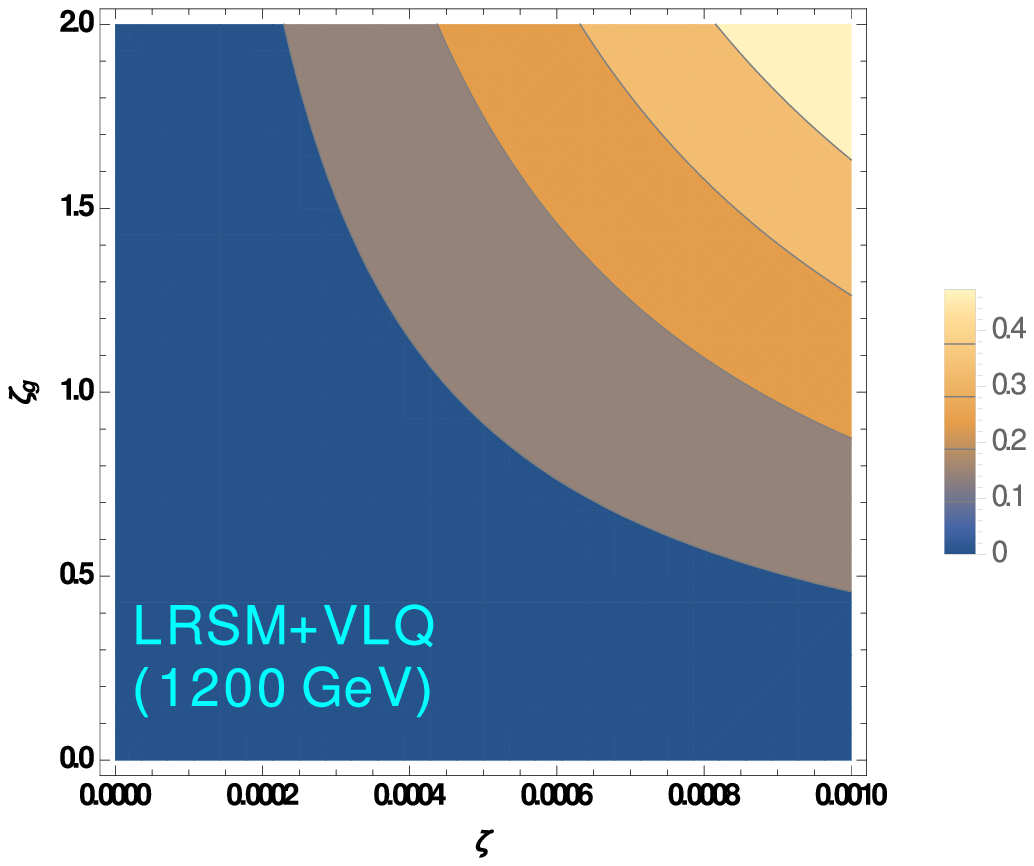}}\\
\subfloat[][]{\includegraphics[width=0.36\textwidth]{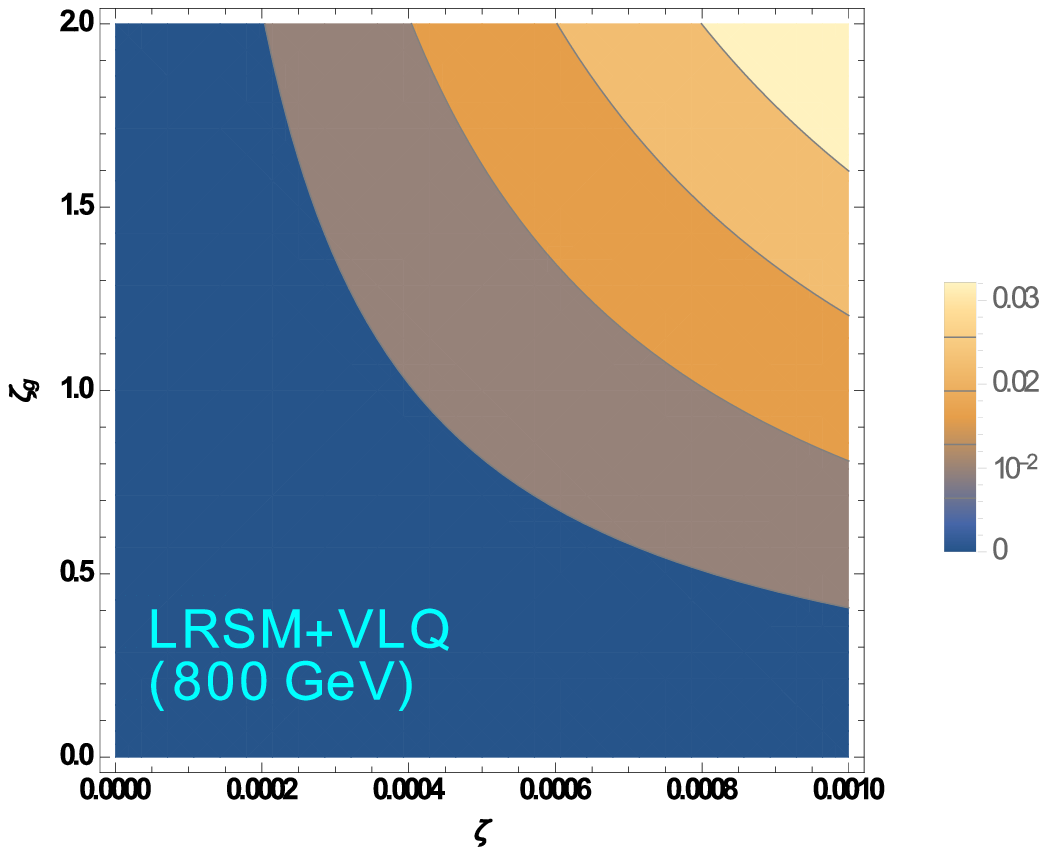}}
\subfloat[][]{\includegraphics[width=0.36\textwidth]{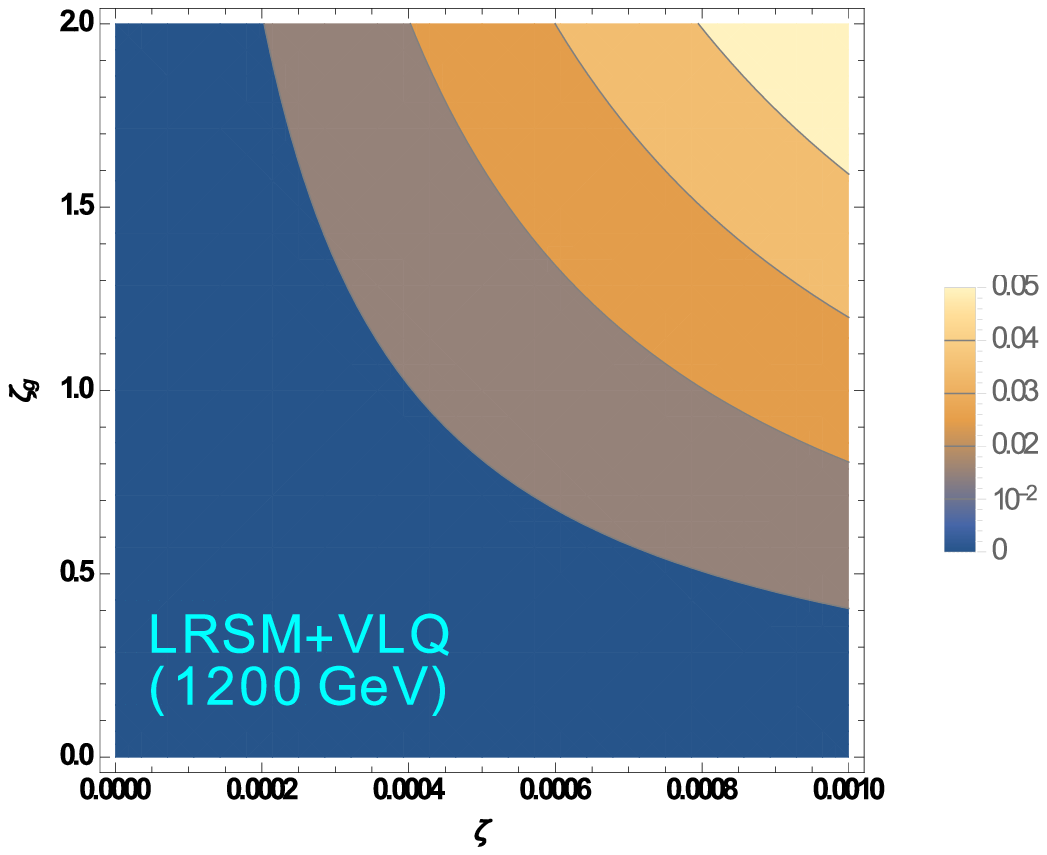}}\\
\caption{\small{Contour plots showing the variation of the polarization $\lambda_\gamma$ as a function of $\zeta$ and $\zeta_g$. The left panels show the plots for LRSM with a VLQ of mass $800$ GeV,
while the right panels display the plots for LRSM with a VLQ of mass $1200$ GeV. The LD amplitudes(in units of GeV$^{-1}$) are $2\times10^{-9}$, $1\times 10^{-8}$, and $8\times10^{-8}$,
for the top, middle and bottom rows respectively. For all the cases, the right-handed CKM elements are set for maximum deviation of the polarization function from its SM value.
The bar-legends for the different contours of $\lambda_\gamma$ are displayed along with the respective figures. Here $0<\zeta<10^{-3}$ and $0<\zeta_g<2$}.}
\label{LD_pol}
\end{figure}
% \end{figure}
As seen in Fig.~\ref{polarization_fig}(a), for very small values of $\zeta$ and $\zeta_{g}$, LRSM approaches the SM and hence in absence of long distance contribution, the polarization is
left handed ($\lambda_{\gamma}=-1$), however as the parameters $\zeta$ and $\zeta_{g}$ increase, the polarization value changes from -1 to +1.
This picture completely changes in the presence of the long distance effects, shown in Fig.~\ref{polarization_fig}(b). Left and right pannels
of Fig.~\ref{LD_pol} show the $\lambda_{\gamma}$ contours for LRSM with an
isosinglet down type vector-like quark of mass $800$ GeV and $1200$ GeV respectively, with LD
amplitudes (in units of GeV$^{-1}$) of $2\times 10^{-9}$ , $1\times 10^{-8}$ and $8\times 10^{-8}$, corresponding to the top, middle and bottom
rows. At the lower end of the range estimated for the LD
amplitude, in
a model with a vector like quark along with LRSM, polarization can be large, even +1 as both $\zeta$ and $\zeta_{g}$ approach their maximum values.
If the LD contributions are larger $\sim 1\times 10^{-8}$ GeV$^{-1}$, the maximum polarization value is $\sim 0.5$, which further reduces to 0.05 for LD amplitude of~$8\times 10^{-8}$ GeV$^{-1}$.

On the experimental side, branching ratios of some of the radiative decays of the $D^0$ meson have been measured by the Belle collaboration~\cite{Belle_rad},
\begin{align*}
&\text{BR}\big(D^{0}\to\rho^{0}\gamma\big)=(1.77\pm 0.3\pm 0.07)\times 10^{-5},\\
&\text{BR}\big(D^{0}\to\phi\gamma\big)=(2.76\pm 0.19\pm 0.10)\times 10^{-5},\\
&\text{BR}\big(D^{0}\to \overline{K}^{*0}\gamma\big)=(4.66\pm 0.21\pm 0.21)\times 10^{-4}.
\end{align*}
If the LD contribution is at its lower limit, then the measured BR($D^{0}\to\rho^{0}\gamma$) can allow some enhancement from the NP SD contribution,
on the other hand, the upper limit of LD saturates the observe BR. The measured BR($D^{0}\to\phi\gamma$) also allows some NP SD contribution.
The upper limit for BR($D^0\rightarrow\omega\gamma$) is $2.4\times10^{-4}$~\cite{pdg} and cannot be saturated by the SM contribution. Also, recently an observation of the photon polarization in the $b\rightarrow s\gamma$ transition was reported by LHCb~\cite{LHCb}.
Photon polarization is obtained by the angular distribution of the photon direction with respect to the plane defined by the momenta of the three final-state hadrons
in their centre of mass frame. A similar technique could be used to measure the photon polarization for the case of $D\rightarrow\omega\gamma$, since the decay
of $\omega$ into three pions will permit the measurement of an up-down asymmetry between the number of events with photons on either side of the plane.
%For the model with a vetor-like quark along with left-right symmetry, the enhancement in the BR for the inclusive $c\to u\gamma$, as well as the value of the photon polarization
%being different from that of the SM, should be reflected in the exclusive modes as well,
%although the results may be less dramatic.
For the model
with left-right symmetry and a vector-like quark, the enhancement in the BR($c\to u\gamma$), as well
as the photon polarization value being different from that of the SM, should be reflected in the
exclusive modes as well, although the results may be weaker.
All the form factors required to estimate the exclusive BR's are neither available from experimental data nor yet extracted from lattice calculations. Hence, we do not attempt to calculate the exact BR's for
specific exclusive modes. Very recently exclusive radiative charm decays have been studied~\cite{Hiller} in heavy quark and hybrid formalism.

\section{Conclusions}\label{V}
Charmed decay modes including radiative ones are expected to be plagued by long distance contributions. For the SM, NLO QCD corrections enhance the short distance $c\to u\gamma$ branching ratio by about $\mathcal{O}(10^{8})$. Further
enhancement of the branching ratio is possible in various new physics models. We show that for certain values of
the parameter space, an enhancement by even up to $\mathcal{O}(10^{10})$ is possible in a left-right symmetric model with a down type vector-like singlet quark at the bare level.
This could be enhanced further by many orders of magnitude after incorporating QCD corrections, enabling the short distance branching fraction to be possibly even
larger than the long distance contribution. Such an enhancement could signal the presence of physics beyond the SM. However, the uncertainty in the size of the long distance contributions,
may not allow this to be easily feasible. Nevertheless measurements of branching ratios of all possible charm radiative modes should be made.
A clearer signature of new physics could be obtained by measurement of the photon polarization, for eg. for the radiative $D\rightarrow\omega\gamma$ mode via a technique similar
to that used recently by LHCb~\cite{LHCb} for the $b\rightarrow s\gamma$ case. We find that for a large region of the parameter space for the vector-like quark
model with left-right symmetry, the photon polarization can be right handed.
% for a left-right symmetric model the enhancement can also be $\mathcal{O}(10^2)$.  depending on the parameters. A model with a vector-like quark along with a left-right symmetry results in a short distance enhancement by $\mathcal{O}(10^3)$. Hence, such models where the enhancement in the branching ratios are much higher than the contribution from the long distance effects, can help in uncovering the presence of such new physics. Hence,  We also suggest a measurement of the photon polarization, for eg. for the radiative $D\rightarrow\omega\gamma$ mode via a technique similar to that used recently by LHCb~\cite{LHCb} for the $b\rightarrow s\gamma$ case as a probe for such NP. 
% that for a considerable region of the parameter space, the polarization is differentconsistent with physics beyond the Standard model. 
For the modes $D\rightarrow K^*\gamma$, $\rho\gamma$ the photon polarization could possibly be determined by looking at the photon conversion to $e^+e^-$~\cite{Grossman}. 
\appendix
\section{The long distance contributions}\label{LDBR}
\begin{table}[ht]
\centering
\caption{The pole-I, pole-II and VMD amplitudes and the exclusive radiative branching ratios.}
\label{br}
\begin{tabular}{|c|ccc|cc|c|}
\hline
 Mode & & $A^{PC}(10^{-8})$ & & $A^{PV}(10^{-8})$ && B.R.\\
      & P-I & P-II & VMD &P-I& VMD & \\                                                
 \hline 
 $D_{s}^{+}\to\rho^{+}\gamma$          & 7.07   &        & $\pm$8.36 &      & $\pm$11.5 & (4.7-13)$\times10^{-4}$\\
 $D_{s}^{+}\to b_{1}^{+}(1235)\gamma$  & 6.19   &        &           &      &           & $4.9\times10^{-5}$ \\
 $D_{s}^{+}\to a_{1}^{+}(1260)\gamma$  & 10.4   &        &           &      &           & $1.4\times10^{-4}$ \\
 $D_{s}^{+}\to a_{2}^{+}(1320)\gamma$  & 18.2   &        &           &      &           & $9.4\times10^{-5}$\\
 $D_{s}^{+}\to K^{*+}\gamma$           & 2.30   &        & $\pm$1.44 &      & $\pm$2.22 & (1.6-5.4)$\times10^{-5}$\\
 $D_{s}^{+}\to K^{*+}(1430)\gamma$     & 5.15   &        &           &      &           & $3.6\times10^{-6}$ \\
 $D_{s}^{+}\to\pi_{2}^{+}(1670)\gamma$ & 7.62   &        &           &      &           & $5.4\times10^{-7}$\\
 $D^{+}\to\rho^{+}\gamma$              & -1.30  & 0.74   & $\pm$1.85 &      & $\pm$2.34 & $(4.2-6.7)\times10^{-5}$\\
 $D^{+}\to K^{*+}\gamma$               &        &        & $\pm$0.46 &      & $\pm$0.6  & $2.7\times10^{-6}$ \\
 $D^{+}\to b_{1}^{+}(1235)\gamma$      & -1.13  &        &           &      &           & $2.4\times10^{-6}$\\
 $D^{+}\to a_{1}^{+}(1260)\gamma$      & -1.91  &        &           &      &           & $6.8\times10^{-6}$\\
 $D^{+}\to a_{2}^{+}(1320)\gamma$      & -3.36  &        &           &      &           & $3.2\times10^{-6}$\\
 $D^{+}\to\pi_{2}^{+}(1670)\gamma$     & -1.40  &        &           &      &           & $5.6\times10^{-9}$\\
 $D^{0}\to\overline{K}^{*0}\gamma$     & -5.21  &        & $\pm$3.62 &      & $\pm$4.77 & (4.6-18)$\times10^{-5}$\\
 $D^{0}\to K_{1}(1270)\gamma$          & -0.016 &        &           &      &           & $1.6\times10^{-10}$\\
 $D^{0}\to K_{1}(1400)\gamma$          & -0.038 &        &           &      &           & $4.7\times10^{-10}$\\
 $D^{0}\to K^{*}(1410)\gamma$          & -0.018 &        &           &      &           & $1\times10^{-10}$\\
 $D^{0}\to\rho^{0}\gamma$              & 1.36   &        & $\pm$1.05 &      & $\pm$1.46 & (5.12-18)$\times10^{-6}$ \\
 $D^{0}\to\omega\gamma$                & -0.703 &        & $\pm$0.897&      & $\pm$1.20 & (3.2-9)$\times10^{-6}$\\
 $D^{0}\to\phi\gamma$                  & 0.318  &        & $\pm$0.956&-0.428& $\pm$1.32 & (4.8-6.4)$\times10^{-6}$\\
 \hline
 \end{tabular}
 \end{table}
In table~\ref{br} we display the results for the calculation of the long-distance $D\rightarrow V\gamma$ amplitudes. The individual numbers for type-I pole and VMD contributions are shown separately along with the branching ratios. These results are essentially an update of the results in Ref.~\cite{Burdman} using the same techniques. The major updates are:
\begin{itemize} 
 \item Inclusion of new modes like $D_s^+\rightarrow\pi_2^+(1670)\gamma$, $D^+\rightarrow\pi_2^+(1670)\gamma$, $D^0\rightarrow K_1(1270)\gamma$, $D^0\rightarrow K_1(1400)\gamma$, $D^0\rightarrow K_1(1410)\gamma$ in the type-I pole amplitudes. 
 \item Updated form factors taken from Ref.~\cite{fajfer_ff} used in calculating the VMD amplitudes.
 \item Updated $V\rightarrow P$ (vector-pseudoscalar) and $T\rightarrow P$ (tensor-pseudoscalar) decay widths from Ref.~\cite{pdg} used for the evaluation of the couplings $h_{V\gamma P}$ and $h_{T\gamma P}$ respectively for the type-I pole amplitudes. 
 \item Inclusion of $\eta-\eta^\prime$ mixing in calculating the type-I $D^0\rightarrow\rho^0\gamma$, $D^0\rightarrow\omega\gamma$ and $D^0\rightarrow\phi\gamma$ amplitudes. The corresponding mixing angles and decay constants have been obtained from Ref.~\cite{eta}.
 \item Inclusion of the parity-violating (PV) part for the $D^{0}\to\phi\gamma$ type-I pole amplitude. The decay constants for the corresponding scalars involved have been taken from Ref.~\cite{f0980} for $f_0(980)$ and \cite{a0980} for $a_0(980)$ respectively.
 \item The decay constants are taken from Ref.~\cite{vector_f} for the light vector mesons and from Ref.~\cite{pdg} for the light pseudoscalar mesons. For the decay constants of the $D^*$ and $D_s^*$ mesons, we use Ref.~\cite{Dst_f}.
\end{itemize}
We have only calculated the type-II pole contribution to the mode $D^+\rightarrow\rho^+\gamma$. This is because:
 \begin{itemize}
  \item The corresponding decay widths for $D_s^*\rightarrow D_s$ and $D^{0^*}\rightarrow D^0$ essential for calculating the type-II pole contributions to the $D_s$ and $D^0$ decay modes respectively are only given as limits in Ref.~\cite{pdg}. 
  \item For decay modes of $D^+$ other than $\rho^+\gamma$ (for eg. $b_1^+(1235)\gamma$, $a_1^+(1260)\gamma$ etc.), the corresponding decay constants for the final state particles are not available. 
  \end{itemize}

\section{The LO and NLO anomalous dimension matrices}\label{LONLOanomalous}
 We provide the effective anomalous dimension matrix in the $\overline{\text{MS}}$ scheme in this appendix. For the case of charm decays we have $q_1=-1/3$, $q_2=2/3$ and $\bar{q}=q_1-q_2$. While evolving the $C_i$'s down from the $M_W$ to the $m_b$ scale, one has to make the assignments $n=3$ and $f=5$. For the corresponding evolution from the $m_b$ to the $m_c$ scale, the values are $n=2$ and $f=4$.
 At the LO level, it is given by
 \begin{equation}\label{g0}
 \hat\gamma_{eff}^0=\left(
\begin{array}{cccccccc}
 -4 & \frac{8}{3} & 0 & -\frac{2}{9} & 0 & 0 & -\frac{4 q_1}{3}-\frac{8 q_2}{81} &
   \frac{173}{162} \\
 12 & 0 & 0 & \frac{4}{3} & 0 & 0 & 8 q_1+\frac{16 q_2}{27} & \frac{70}{27} \\
 0 & 0 & 0 & -\frac{52}{3} & 0 & 2 & \frac{176 q_2}{27} & \frac{14}{27} \\
 0 & 0 & -\frac{40}{9} & \frac{4 f}{3}-\frac{160}{9} & \frac{4}{9} & \frac{5}{6} &
   \left(\frac{16 f}{27}-\frac{88}{81}\right) q_2 & \frac{74}{81}-\frac{49 f}{54} \\
 0 & 0 & 0 & -\frac{256}{3} & 0 & 20 & \frac{6272 q_2}{27} & 36 f+\frac{1736}{27} \\
 0 & 0 & -\frac{256}{9} & \frac{40 f}{3}-\frac{544}{9} & \frac{40}{9} & -\frac{2}{3} &
   48 n \bar{q}+\left(\frac{1456 f}{27}-\frac{3136}{81}\right) q_2 & \frac{160
   f}{27}+\frac{2372}{81} \\
 0 & 0 & 0 & 0 & 0 & 0 & \frac{32}{3} & 0 \\
 0 & 0 & 0 & 0 & 0 & 0 & \frac{32 q_2}{3} & \frac{28}{3} 
\end{array}
\right).
\end{equation}
At the NLO level, due to its large size, we present the matrix in $8\times6$ and $8\times2$ blocks. It reads
\begin{equation}\label{g11}
\hat\gamma_{eff}^{1^{8\times6}}=\left(
\begin{array}{cccccc}
 \frac{16 f}{9}-\frac{145}{3} & \frac{40 f}{27}-26 & -\frac{1412}{243} & -\frac{1369}{243} & \frac{134}{243} & -\frac{35}{162} \\
 \frac{20 f}{3}-45 & -\frac{28}{3} & -\frac{416}{81} & \frac{1280}{81} & \frac{56}{81} & \frac{35}{27}  \\
 0 & 0 & -\frac{4468}{81} & -\frac{52 f}{9}-\frac{29129}{81} & \frac{400}{81} & \frac{3493}{108}-\frac{2 f}{9}  \\
 0 & 0 & \frac{368 f}{81}-\frac{13678}{243} & \frac{1334 f}{81}-\frac{79409}{243} & \frac{509}{486}-\frac{8 f}{81} & \frac{13499}{648}-\frac{5 f}{27}\\
 0 & 0 & -\frac{160 f}{9}-\frac{244480}{81} & -\frac{2200 f}{9}-\frac{29648}{81} & \frac{16 f}{9}+\frac{23116}{81} & \frac{148 f}{9}+\frac{3886}{27} \\
 0 & 0 & \frac{77600}{243}-\frac{1264 f}{81} & \frac{164 f}{81}-\frac{28808}{243} & \frac{400 f}{81}-\frac{20324}{243} & \frac{622 f}{27}-\frac{21211}{162}\\
 0 & 0 & 0 & 0 & 0 & 0 \\
 0 & 0 & 0 & 0 & 0 & 0 
\end{array}
\right)
\end{equation}
\begin{equation}\label{g12}
\hat\gamma_{eff}^{1^{8\times2}}=\left(
\begin{array}{cc}
 \left(\frac{2 f}{27}-\frac{374}{27}\right) q_1+\left(\frac{64 f}{729}-\frac{12614}{729}\right) q_2 & \frac{431 f}{5832}+\frac{65867}{5832} \\
 \left(\frac{136}{9}-\frac{4 f}{9}\right) q_1+\left(\frac{2332}{243}-\frac{128 f}{243}\right) q_2 & \frac{10577}{486}-\frac{917 f}{972} \\
 -\frac{112 n \bar{q}}{3}-\left(\frac{4720 f}{243}+\frac{97876}{243}\right) q_2 & \frac{42524}{243}-\frac{2398 f}{243} \\
 \left(-\frac{32 f^2}{243}+\frac{4448 f}{729}+\frac{70376}{729}\right) q_2-\frac{140 n \bar{q}}{9} & -\frac{253 f^2}{486}-\frac{39719 f}{5832}-\frac{159718}{729} \\
 -\frac{3136 n \bar{q}}{3}-\left(\frac{188608 f}{243}+\frac{1764752}{243}\right) q_2 & -14 f^2+\frac{140954 f}{243}+\frac{2281576}{243} \\
 \left(-\frac{56 f}{3}-\frac{1136}{9}\right) n \bar{q}-\left(\frac{5432 f^2}{243}-\frac{232112 f}{729}+\frac{4193840}{729}\right) q_2 & -\frac{6031 f^2}{486}-\frac{15431
   f}{1458}-\frac{3031517}{729} \\
 \frac{1936}{9}-\frac{224 f}{27} & 0 \\
 \left(\frac{368}{3}-\frac{224 f}{27}\right) q_2 & \frac{1456}{9}-\frac{61 f}{27} 
\end{array}
\right)
\end{equation}

\section{The functions $\textbf{G}_{\textbf{p}}$}\label{G}
In this appendix, we provide the form of the $G_p$ functions mentioned in sections~\ref{vector_like} taken from Ref.~\cite{Kim}.
\begin{equation}
 G_p(r)=\int_0^1dx\int_0^{1-x}dy\frac{1}{\Lambda_p(r,x,y)}g_p(r,x,y)
\end{equation}
where $p=1,2$ and
\begin{eqnarray}
 \Lambda_1(r,x,y)&=&1-x+rx-r_ixy-r_fxz\nonumber\\
 \Lambda_2(r,x,y)&=&x+r(1-x)-r_ixy-r_fxz\nonumber\\
 g_1(r,x,y)&=&1-x+z+y(1-2x)+rx(1-y)-r_fxz\nonumber\\
 g_2(r,x,y)&=&-2x(1-y)+r(-1+x+xy)+r_fxz.
\end{eqnarray}
In the above, $r_i=m_c^2/M_W^2$, $r_f=m_u^2/M_W^2$ and $r=m^2/M_W^2$ with $m_c$, $m_u$ and $M_W$ denoting the masses of the charm quark, up quark and the W-boson respectively. The $m$ stands for the masses of the down-type quarks running in the fermionic penguin loop for a $c\rightarrow u\gamma$ transition. Hence, for the SM $m=(m_d,m_s,m_b)$ and for the vector-like quark model $m=(m_d,m_s,m_b,m_{b^\prime})$ with $m_d$, $m_s$, $m_b$ and $m_{b^\prime}$ standing for the masses of the $d$, $s$, $b$ and $b^{\prime}$ quarks respectively.

\acknowledgments

NS thanks R.Sinha for discussions. The authors thank Sandip Pakvasa for discussions and comments. They also thank Jim Libby for his suggestions. Gudrun Hiller and Stefan de Boer are also thanked for their comments.
% 
%\bibliography{references}

\begin{thebibliography}{47}%
\makeatletter
\providecommand \@ifxundefined [1]{%
 \@ifx{#1\undefined}
}%
\providecommand \@ifnum [1]{%
 \ifnum #1\expandafter \@firstoftwo
 \else \expandafter \@secondoftwo
 \fi
}%
\providecommand \@ifx [1]{%
 \ifx #1\expandafter \@firstoftwo
 \else \expandafter \@secondoftwo
 \fi
}%
\providecommand \natexlab [1]{#1}%
\providecommand \enquote  [1]{``#1''}%
\providecommand \bibnamefont  [1]{#1}%
\providecommand \bibfnamefont [1]{#1}%
\providecommand \citenamefont [1]{#1}%
\providecommand \href@noop [0]{\@secondoftwo}%
\providecommand \href [0]{\begingroup \@sanitize@url \@href}%
\providecommand \@href[1]{\@@startlink{#1}\@@href}%
\providecommand \@@href[1]{\endgroup#1\@@endlink}%
\providecommand \@sanitize@url [0]{\catcode `\\12\catcode `\$12\catcode
  `\&12\catcode `\#12\catcode `\^12\catcode `\_12\catcode `\%12\relax}%
\providecommand \@@startlink[1]{}%
\providecommand \@@endlink[0]{}%
\providecommand \url  [0]{\begingroup\@sanitize@url \@url }%
\providecommand \@url [1]{\endgroup\@href {#1}{\urlprefix }}%
\providecommand \urlprefix  [0]{URL }%
\providecommand \Eprint [0]{\href }%
\providecommand \doibase [0]{http://dx.doi.org/}%
\providecommand \selectlanguage [0]{\@gobble}%
\providecommand \bibinfo  [0]{\@secondoftwo}%
\providecommand \bibfield  [0]{\@secondoftwo}%
\providecommand \translation [1]{[#1]}%
\providecommand \BibitemOpen [0]{}%
\providecommand \bibitemStop [0]{}%
\providecommand \bibitemNoStop [0]{.\EOS\space}%
\providecommand \EOS [0]{\spacefactor3000\relax}%
\providecommand \BibitemShut  [1]{\csname bibitem#1\endcsname}%
\let\auto@bib@innerbib\@empty
%</preamble>
\bibitem [{\citenamefont {Ligeti}(2016)}]{Ligeti}%
  \BibitemOpen
  \bibfield  {author} {\bibinfo {author} {\bibfnamefont {Z.}~\bibnamefont
  {Ligeti}},\ }\bibfield  {booktitle} {\emph {\bibinfo {booktitle}
  {{Proceedings, 27th International Symposium on Lepton Photon Interactions at
  High Energy (LP15): Ljubljana, Slovenia, August 17-22, 2015}}},\ }\href@noop
  {} {\bibfield  {journal} {\bibinfo  {journal} {PoS}\ }\textbf {\bibinfo
  {volume} {LeptonPhoton2015}},\ \bibinfo {pages} {031} (\bibinfo {year}
  {2016})},\ \Eprint {http://arxiv.org/abs/1606.02756} {arXiv:1606.02756
  [hep-ph]} \BibitemShut {NoStop}%
%%CITATION = ARXIV:1606.02756;%%
\bibitem [{\citenamefont {Burdman}\ \emph {et~al.}(2002)\citenamefont
  {Burdman}, \citenamefont {Golowich}, \citenamefont {Hewett},\ and\
  \citenamefont {Pakvasa}}]{pakvasa_2002}%
  \BibitemOpen
  \bibfield  {author} {\bibinfo {author} {\bibfnamefont {G.}~\bibnamefont
  {Burdman}}, \bibinfo {author} {\bibfnamefont {E.}~\bibnamefont {Golowich}},
  \bibinfo {author} {\bibfnamefont {J.~L.}\ \bibnamefont {Hewett}}, \ and\
  \bibinfo {author} {\bibfnamefont {S.}~\bibnamefont {Pakvasa}},\ }\href
  {\doibase 10.1103/PhysRevD.66.014009} {\bibfield  {journal} {\bibinfo
  {journal} {Phys. Rev.}\ }\textbf {\bibinfo {volume} {D66}},\ \bibinfo {pages}
  {014009} (\bibinfo {year} {2002})},\ \Eprint
  {http://arxiv.org/abs/hep-ph/0112235} {arXiv:hep-ph/0112235 [hep-ph]}
  \BibitemShut {NoStop}%
%%CITATION = HEP-PH/0112235;%%
\bibitem [{\citenamefont {Fajfer}(2015)}]{fajfer_rare}%
  \BibitemOpen
  \bibfield  {author} {\bibinfo {author} {\bibfnamefont {S.}~\bibnamefont
  {Fajfer}},\ }in\ \href
  {http://inspirehep.net/record/1392013/files/arXiv:1509.01997.pdf} {\emph
  {\bibinfo {booktitle} {{7th International Workshop on Charm Physics (Charm
  2015) Detroit, MI, USA, May 18-22, 2015}}}}\ (\bibinfo {year} {2015})\
  \Eprint {http://arxiv.org/abs/1509.01997} {arXiv:1509.01997 [hep-ph]}
  \BibitemShut {NoStop}%
%%CITATION = ARXIV:1509.01997;%%
\bibitem [{\citenamefont {Petrov}(2016)}]{petrov_rare}%
  \BibitemOpen
  \bibfield  {author} {\bibinfo {author} {\bibfnamefont {A.~A.}\ \bibnamefont
  {Petrov}},\ }\bibfield  {booktitle} {\emph {\bibinfo {booktitle}
  {{Proceedings, 16th International Conference on B-Physics at Frontier
  Machines (Beauty 2016): Marseille, France, May 2-6, 2016}}},\ }\href@noop {}
  {\bibfield  {journal} {\bibinfo  {journal} {PoS}\ }\textbf {\bibinfo {volume}
  {BEAUTY2016}},\ \bibinfo {pages} {011} (\bibinfo {year} {2016})},\ \Eprint
  {http://arxiv.org/abs/1609.04448} {arXiv:1609.04448 [hep-ph]} \BibitemShut
  {NoStop}%
%%CITATION = ARXIV:1609.04448;%%
\bibitem [{\citenamefont {de~Boer}\ and\ \citenamefont {Hiller}(2016)}]{Boer}%
  \BibitemOpen
  \bibfield  {author} {\bibinfo {author} {\bibfnamefont {S.}~\bibnamefont
  {de~Boer}}\ and\ \bibinfo {author} {\bibfnamefont {G.}~\bibnamefont
  {Hiller}},\ }\href {\doibase 10.1103/PhysRevD.93.074001} {\bibfield
  {journal} {\bibinfo  {journal} {Phys. Rev.}\ }\textbf {\bibinfo {volume}
  {D93}},\ \bibinfo {pages} {074001} (\bibinfo {year} {2016})},\ \Eprint
  {http://arxiv.org/abs/1510.00311} {arXiv:1510.00311 [hep-ph]} \BibitemShut
  {NoStop}%
%%CITATION = ARXIV:1510.00311;%%
\bibitem [{\citenamefont {Fajfer}\ \emph {et~al.}(2000)\citenamefont {Fajfer},
  \citenamefont {Prelovsek}, \citenamefont {Singer},\ and\ \citenamefont
  {Wyler}}]{FajferRatio}%
  \BibitemOpen
  \bibfield  {author} {\bibinfo {author} {\bibfnamefont {S.}~\bibnamefont
  {Fajfer}}, \bibinfo {author} {\bibfnamefont {S.}~\bibnamefont {Prelovsek}},
  \bibinfo {author} {\bibfnamefont {P.}~\bibnamefont {Singer}}, \ and\ \bibinfo
  {author} {\bibfnamefont {D.}~\bibnamefont {Wyler}},\ }\href {\doibase
  10.1016/S0370-2693(00)00731-0} {\bibfield  {journal} {\bibinfo  {journal}
  {Phys. Lett.}\ }\textbf {\bibinfo {volume} {B487}},\ \bibinfo {pages} {81}
  (\bibinfo {year} {2000})},\ \Eprint {http://arxiv.org/abs/hep-ph/0006054}
  {arXiv:hep-ph/0006054 [hep-ph]} \BibitemShut {NoStop}%
%%CITATION = HEP-PH/0006054;%%
\bibitem [{\citenamefont {Babu}\ \emph {et~al.}(1988)\citenamefont {Babu},
  \citenamefont {He}, \citenamefont {Li},\ and\ \citenamefont
  {Pakvasa}}]{BabuHeLiPakvasa}%
  \BibitemOpen
  \bibfield  {author} {\bibinfo {author} {\bibfnamefont {K.~S.}\ \bibnamefont
  {Babu}}, \bibinfo {author} {\bibfnamefont {X.~G.}\ \bibnamefont {He}},
  \bibinfo {author} {\bibfnamefont {X.}~\bibnamefont {Li}}, \ and\ \bibinfo
  {author} {\bibfnamefont {S.}~\bibnamefont {Pakvasa}},\ }\href {\doibase
  10.1016/0370-2693(88)90994-X} {\bibfield  {journal} {\bibinfo  {journal}
  {Phys. Lett.}\ }\textbf {\bibinfo {volume} {B205}},\ \bibinfo {pages} {540}
  (\bibinfo {year} {1988})}\BibitemShut {NoStop}%
%%CITATION = PHLTA,B205,540;%%
\bibitem [{\citenamefont {Cacciapaglia}\ \emph {et~al.}(2012)\citenamefont
  {Cacciapaglia}, \citenamefont {Deandrea}, \citenamefont {Panizzi},
  \citenamefont {Gaur}, \citenamefont {Harada},\ and\ \citenamefont
  {Okada}}]{Cacciapaglia2011}%
  \BibitemOpen
  \bibfield  {author} {\bibinfo {author} {\bibfnamefont {G.}~\bibnamefont
  {Cacciapaglia}}, \bibinfo {author} {\bibfnamefont {A.}~\bibnamefont
  {Deandrea}}, \bibinfo {author} {\bibfnamefont {L.}~\bibnamefont {Panizzi}},
  \bibinfo {author} {\bibfnamefont {N.}~\bibnamefont {Gaur}}, \bibinfo {author}
  {\bibfnamefont {D.}~\bibnamefont {Harada}}, \ and\ \bibinfo {author}
  {\bibfnamefont {Y.}~\bibnamefont {Okada}},\ }\href {\doibase
  10.1007/JHEP03(2012)070} {\bibfield  {journal} {\bibinfo  {journal} {JHEP}\
  }\textbf {\bibinfo {volume} {03}},\ \bibinfo {pages} {070} (\bibinfo {year}
  {2012})},\ \Eprint {http://arxiv.org/abs/1108.6329} {arXiv:1108.6329
  [hep-ph]} \BibitemShut {NoStop}%
%%CITATION = ARXIV:1108.6329;%%
\bibitem [{\citenamefont {Botella}\ \emph {et~al.}(2013)\citenamefont
  {Botella}, \citenamefont {Nebot},\ and\ \citenamefont
  {Branco}}]{Botella2013}%
  \BibitemOpen
  \bibfield  {author} {\bibinfo {author} {\bibfnamefont {F.~J.}\ \bibnamefont
  {Botella}}, \bibinfo {author} {\bibfnamefont {M.}~\bibnamefont {Nebot}}, \
  and\ \bibinfo {author} {\bibfnamefont {G.~C.}\ \bibnamefont {Branco}},\
  }\bibfield  {booktitle} {\emph {\bibinfo {booktitle} {{Proceedings, 3rd
  Symposium on Prospects in the Physics of Discrete Symmetries (DISCRETE 2012):
  Lisbon, Portugal, December 3-7, 2012}}},\ }\href {\doibase
  10.1088/1742-6596/447/1/012061} {\bibfield  {journal} {\bibinfo  {journal}
  {J. Phys. Conf. Ser.}\ }\textbf {\bibinfo {volume} {447}},\ \bibinfo {pages}
  {012061} (\bibinfo {year} {2013})}\BibitemShut {NoStop}%
%%CITATION = 00462,447,012061;%%
\bibitem [{\citenamefont {Ishiwata}\ \emph {et~al.}(2015)\citenamefont
  {Ishiwata}, \citenamefont {Ligeti},\ and\ \citenamefont
  {Wise}}]{Ishiwata2015}%
  \BibitemOpen
  \bibfield  {author} {\bibinfo {author} {\bibfnamefont {K.}~\bibnamefont
  {Ishiwata}}, \bibinfo {author} {\bibfnamefont {Z.}~\bibnamefont {Ligeti}}, \
  and\ \bibinfo {author} {\bibfnamefont {M.~B.}\ \bibnamefont {Wise}},\ }\href
  {\doibase 10.1007/JHEP10(2015)027} {\bibfield  {journal} {\bibinfo  {journal}
  {JHEP}\ }\textbf {\bibinfo {volume} {10}},\ \bibinfo {pages} {027} (\bibinfo
  {year} {2015})},\ \Eprint {http://arxiv.org/abs/1506.03484} {arXiv:1506.03484
  [hep-ph]} \BibitemShut {NoStop}%
%%CITATION = ARXIV:1506.03484;%%
\bibitem [{\citenamefont {Bobeth}\ \emph {et~al.}(2016)\citenamefont {Bobeth},
  \citenamefont {Buras}, \citenamefont {Celis},\ and\ \citenamefont
  {Jung}}]{Bobeth2016}%
  \BibitemOpen
  \bibfield  {author} {\bibinfo {author} {\bibfnamefont {C.}~\bibnamefont
  {Bobeth}}, \bibinfo {author} {\bibfnamefont {A.~J.}\ \bibnamefont {Buras}},
  \bibinfo {author} {\bibfnamefont {A.}~\bibnamefont {Celis}}, \ and\ \bibinfo
  {author} {\bibfnamefont {M.}~\bibnamefont {Jung}},\ }\href@noop {} {\
  (\bibinfo {year} {2016})},\ \Eprint {http://arxiv.org/abs/1609.04783}
  {arXiv:1609.04783 [hep-ph]} \BibitemShut {NoStop}%
%%CITATION = ARXIV:1609.04783;%%
\bibitem [{\citenamefont {Alok}\ \emph {et~al.}(2016)\citenamefont {Alok},
  \citenamefont {Banerjee}, \citenamefont {Kumar},\ and\ \citenamefont
  {Sankar}}]{uma}%
  \BibitemOpen
  \bibfield  {author} {\bibinfo {author} {\bibfnamefont {A.~K.}\ \bibnamefont
  {Alok}}, \bibinfo {author} {\bibfnamefont {S.}~\bibnamefont {Banerjee}},
  \bibinfo {author} {\bibfnamefont {D.}~\bibnamefont {Kumar}}, \ and\ \bibinfo
  {author} {\bibfnamefont {S.~U.}\ \bibnamefont {Sankar}},\ }\href {\doibase
  http://dx.doi.org/10.1016/j.nuclphysb.2016.03.012} {\bibfield  {journal}
  {\bibinfo  {journal} {Nuclear Physics B}\ }\textbf {\bibinfo {volume}
  {906}},\ \bibinfo {pages} {321 } (\bibinfo {year} {2016})}\BibitemShut
  {NoStop}%
\bibitem [{\citenamefont {Atwood}\ \emph {et~al.}(2007)\citenamefont {Atwood},
  \citenamefont {Gershon}, \citenamefont {Hazumi},\ and\ \citenamefont
  {Soni}}]{Atwood}%
  \BibitemOpen
  \bibfield  {author} {\bibinfo {author} {\bibfnamefont {D.}~\bibnamefont
  {Atwood}}, \bibinfo {author} {\bibfnamefont {T.}~\bibnamefont {Gershon}},
  \bibinfo {author} {\bibfnamefont {M.}~\bibnamefont {Hazumi}}, \ and\ \bibinfo
  {author} {\bibfnamefont {A.}~\bibnamefont {Soni}},\ }\href@noop {} {\
  (\bibinfo {year} {2007})},\ \Eprint {http://arxiv.org/abs/hep-ph/0701021}
  {arXiv:hep-ph/0701021 [hep-ph]} \BibitemShut {NoStop}%
%%CITATION = HEP-PH/0701021;%%
\bibitem [{\citenamefont {Gronau}\ and\ \citenamefont {Pirjol}(2002)}]{Gronau}%
  \BibitemOpen
  \bibfield  {author} {\bibinfo {author} {\bibfnamefont {M.}~\bibnamefont
  {Gronau}}\ and\ \bibinfo {author} {\bibfnamefont {D.}~\bibnamefont
  {Pirjol}},\ }\href {\doibase 10.1103/PhysRevD.66.054008} {\bibfield
  {journal} {\bibinfo  {journal} {Phys. Rev.}\ }\textbf {\bibinfo {volume}
  {D66}},\ \bibinfo {pages} {054008} (\bibinfo {year} {2002})},\ \Eprint
  {http://arxiv.org/abs/hep-ph/0205065} {arXiv:hep-ph/0205065 [hep-ph]}
  \BibitemShut {NoStop}%
%%CITATION = HEP-PH/0205065;%%
\bibitem [{\citenamefont {Burdman}\ \emph {et~al.}(1995)\citenamefont
  {Burdman}, \citenamefont {Golowich}, \citenamefont {Hewett},\ and\
  \citenamefont {Pakvasa}}]{Burdman}%
  \BibitemOpen
  \bibfield  {author} {\bibinfo {author} {\bibfnamefont {G.}~\bibnamefont
  {Burdman}}, \bibinfo {author} {\bibfnamefont {E.}~\bibnamefont {Golowich}},
  \bibinfo {author} {\bibfnamefont {J.~L.}\ \bibnamefont {Hewett}}, \ and\
  \bibinfo {author} {\bibfnamefont {S.}~\bibnamefont {Pakvasa}},\ }\href
  {\doibase 10.1103/PhysRevD.52.6383} {\bibfield  {journal} {\bibinfo
  {journal} {Phys. Rev.}\ }\textbf {\bibinfo {volume} {D52}},\ \bibinfo {pages}
  {6383} (\bibinfo {year} {1995})},\ \Eprint
  {http://arxiv.org/abs/hep-ph/9502329} {arXiv:hep-ph/9502329 [hep-ph]}
  \BibitemShut {NoStop}%
%%CITATION = HEP-PH/9502329;%%
\bibitem [{\citenamefont {Inami}\ and\ \citenamefont {Lim}(1981)}]{InamiLim}%
  \BibitemOpen
  \bibfield  {author} {\bibinfo {author} {\bibfnamefont {T.}~\bibnamefont
  {Inami}}\ and\ \bibinfo {author} {\bibfnamefont {C.~S.}\ \bibnamefont
  {Lim}},\ }\href {\doibase 10.1143/PTP.65.297} {\bibfield  {journal} {\bibinfo
   {journal} {Prog. Theor. Phys.}\ }\textbf {\bibinfo {volume} {65}},\ \bibinfo
  {pages} {297} (\bibinfo {year} {1981})},\ \bibinfo {note} {[Erratum: Prog.
  Theor. Phys.65,1772(1981)]}\BibitemShut {NoStop}%
%%CITATION = PTPKA,65,297;%%
\bibitem [{\citenamefont {Ho-Kim}\ and\ \citenamefont {Pham}(2000)}]{Kim}%
  \BibitemOpen
  \bibfield  {author} {\bibinfo {author} {\bibfnamefont {Q.}~\bibnamefont
  {Ho-Kim}}\ and\ \bibinfo {author} {\bibfnamefont {X.-Y.}\ \bibnamefont
  {Pham}},\ }\href {\doibase 10.1103/PhysRevD.61.013008} {\bibfield  {journal}
  {\bibinfo  {journal} {Phys. Rev.}\ }\textbf {\bibinfo {volume} {D61}},\
  \bibinfo {pages} {013008} (\bibinfo {year} {2000})},\ \Eprint
  {http://arxiv.org/abs/hep-ph/9906235} {arXiv:hep-ph/9906235 [hep-ph]}
  \BibitemShut {NoStop}%
%%CITATION = HEP-PH/9906235;%%
\bibitem [{\citenamefont {Greub}\ \emph {et~al.}(1996)\citenamefont {Greub},
  \citenamefont {Hurth}, \citenamefont {Misiak},\ and\ \citenamefont
  {Wyler}}]{greub}%
  \BibitemOpen
  \bibfield  {author} {\bibinfo {author} {\bibfnamefont {C.}~\bibnamefont
  {Greub}}, \bibinfo {author} {\bibfnamefont {T.}~\bibnamefont {Hurth}},
  \bibinfo {author} {\bibfnamefont {M.}~\bibnamefont {Misiak}}, \ and\ \bibinfo
  {author} {\bibfnamefont {D.}~\bibnamefont {Wyler}},\ }\href {\doibase
  10.1016/0370-2693(96)00694-6} {\bibfield  {journal} {\bibinfo  {journal}
  {Phys. Lett.}\ }\textbf {\bibinfo {volume} {B382}},\ \bibinfo {pages} {415}
  (\bibinfo {year} {1996})},\ \Eprint {http://arxiv.org/abs/hep-ph/9603417}
  {arXiv:hep-ph/9603417 [hep-ph]} \BibitemShut {NoStop}%
%%CITATION = HEP-PH/9603417;%%
\bibitem [{\citenamefont {Fajfer}\ \emph {et~al.}(2003)\citenamefont {Fajfer},
  \citenamefont {Singer},\ and\ \citenamefont {Zupan}}]{fajfer_wc}%
  \BibitemOpen
  \bibfield  {author} {\bibinfo {author} {\bibfnamefont {S.}~\bibnamefont
  {Fajfer}}, \bibinfo {author} {\bibfnamefont {P.}~\bibnamefont {Singer}}, \
  and\ \bibinfo {author} {\bibfnamefont {J.}~\bibnamefont {Zupan}},\ }\href
  {\doibase 10.1140/epjc/s2002-01090-5} {\bibfield  {journal} {\bibinfo
  {journal} {Eur. Phys. J.}\ }\textbf {\bibinfo {volume} {C27}},\ \bibinfo
  {pages} {201} (\bibinfo {year} {2003})},\ \Eprint
  {http://arxiv.org/abs/hep-ph/0209250} {arXiv:hep-ph/0209250 [hep-ph]}
  \BibitemShut {NoStop}%
%%CITATION = HEP-PH/0209250;%%
\bibitem [{\citenamefont {de~Boer}\ \emph {et~al.}(2016)\citenamefont
  {de~Boer}, \citenamefont {Müller},\ and\ \citenamefont
  {Seidel}}]{muller_wc}%
  \BibitemOpen
  \bibfield  {author} {\bibinfo {author} {\bibfnamefont {S.}~\bibnamefont
  {de~Boer}}, \bibinfo {author} {\bibfnamefont {B.}~\bibnamefont {Müller}}, \
  and\ \bibinfo {author} {\bibfnamefont {D.}~\bibnamefont {Seidel}},\ }\href
  {\doibase 10.1007/JHEP08(2016)091} {\bibfield  {journal} {\bibinfo  {journal}
  {JHEP}\ }\textbf {\bibinfo {volume} {08}},\ \bibinfo {pages} {091} (\bibinfo
  {year} {2016})},\ \Eprint {http://arxiv.org/abs/1606.05521} {arXiv:1606.05521
  [hep-ph]} \BibitemShut {NoStop}%
%%CITATION = ARXIV:1606.05521;%%
\bibitem [{\citenamefont {Buchalla}\ \emph {et~al.}(1996)\citenamefont
  {Buchalla}, \citenamefont {Buras},\ and\ \citenamefont
  {Lautenbacher}}]{Buchalla_Buras}%
  \BibitemOpen
  \bibfield  {author} {\bibinfo {author} {\bibfnamefont {G.}~\bibnamefont
  {Buchalla}}, \bibinfo {author} {\bibfnamefont {A.~J.}\ \bibnamefont {Buras}},
  \ and\ \bibinfo {author} {\bibfnamefont {M.~E.}\ \bibnamefont
  {Lautenbacher}},\ }\href {\doibase 10.1103/RevModPhys.68.1125} {\bibfield
  {journal} {\bibinfo  {journal} {Rev. Mod. Phys.}\ }\textbf {\bibinfo {volume}
  {68}},\ \bibinfo {pages} {1125} (\bibinfo {year} {1996})},\ \Eprint
  {http://arxiv.org/abs/hep-ph/9512380} {arXiv:hep-ph/9512380 [hep-ph]}
  \BibitemShut {NoStop}%
%%CITATION = HEP-PH/9512380;%%
\bibitem [{\citenamefont {Beneke}\ \emph {et~al.}(2001)\citenamefont {Beneke},
  \citenamefont {Feldmann},\ and\ \citenamefont {Seidel}}]{ref_12_muller_wc}%
  \BibitemOpen
  \bibfield  {author} {\bibinfo {author} {\bibfnamefont {M.}~\bibnamefont
  {Beneke}}, \bibinfo {author} {\bibfnamefont {T.}~\bibnamefont {Feldmann}}, \
  and\ \bibinfo {author} {\bibfnamefont {D.}~\bibnamefont {Seidel}},\ }\href
  {\doibase 10.1016/S0550-3213(01)00366-2} {\bibfield  {journal} {\bibinfo
  {journal} {Nucl. Phys.}\ }\textbf {\bibinfo {volume} {B612}},\ \bibinfo
  {pages} {25} (\bibinfo {year} {2001})},\ \Eprint
  {http://arxiv.org/abs/hep-ph/0106067} {arXiv:hep-ph/0106067 [hep-ph]}
  \BibitemShut {NoStop}%
%%CITATION = HEP-PH/0106067;%%
\bibitem [{\citenamefont {Chetyrkin}\ \emph {et~al.}(1998)\citenamefont
  {Chetyrkin}, \citenamefont {Misiak},\ and\ \citenamefont
  {Munz}}]{ref_7_muller_wc}%
  \BibitemOpen
  \bibfield  {author} {\bibinfo {author} {\bibfnamefont {K.~G.}\ \bibnamefont
  {Chetyrkin}}, \bibinfo {author} {\bibfnamefont {M.}~\bibnamefont {Misiak}}, \
  and\ \bibinfo {author} {\bibfnamefont {M.}~\bibnamefont {Munz}},\ }\href
  {\doibase 10.1016/S0550-3213(98)00131-X} {\bibfield  {journal} {\bibinfo
  {journal} {Nucl. Phys.}\ }\textbf {\bibinfo {volume} {B520}},\ \bibinfo
  {pages} {279} (\bibinfo {year} {1998})},\ \Eprint
  {http://arxiv.org/abs/hep-ph/9711280} {arXiv:hep-ph/9711280 [hep-ph]}
  \BibitemShut {NoStop}%
%%CITATION = HEP-PH/9711280;%%
\bibitem [{\citenamefont {Gorbahn}\ and\ \citenamefont
  {Haisch}(2005)}]{ref_10_muller_wc}%
  \BibitemOpen
  \bibfield  {author} {\bibinfo {author} {\bibfnamefont {M.}~\bibnamefont
  {Gorbahn}}\ and\ \bibinfo {author} {\bibfnamefont {U.}~\bibnamefont
  {Haisch}},\ }\href {\doibase 10.1016/j.nuclphysb.2005.01.047} {\bibfield
  {journal} {\bibinfo  {journal} {Nucl. Phys.}\ }\textbf {\bibinfo {volume}
  {B713}},\ \bibinfo {pages} {291} (\bibinfo {year} {2005})},\ \Eprint
  {http://arxiv.org/abs/hep-ph/0411071} {arXiv:hep-ph/0411071 [hep-ph]}
  \BibitemShut {NoStop}%
%%CITATION = HEP-PH/0411071;%%
\bibitem [{\citenamefont {Gorbahn}\ \emph {et~al.}(2005)\citenamefont
  {Gorbahn}, \citenamefont {Haisch},\ and\ \citenamefont
  {Misiak}}]{ref_11_muller_wc}%
  \BibitemOpen
  \bibfield  {author} {\bibinfo {author} {\bibfnamefont {M.}~\bibnamefont
  {Gorbahn}}, \bibinfo {author} {\bibfnamefont {U.}~\bibnamefont {Haisch}}, \
  and\ \bibinfo {author} {\bibfnamefont {M.}~\bibnamefont {Misiak}},\ }\href
  {\doibase 10.1103/PhysRevLett.95.102004} {\bibfield  {journal} {\bibinfo
  {journal} {Phys. Rev. Lett.}\ }\textbf {\bibinfo {volume} {95}},\ \bibinfo
  {pages} {102004} (\bibinfo {year} {2005})},\ \Eprint
  {http://arxiv.org/abs/hep-ph/0504194} {arXiv:hep-ph/0504194 [hep-ph]}
  \BibitemShut {NoStop}%
%%CITATION = HEP-PH/0504194;%%
\bibitem [{\citenamefont {Chetyrkin}\ \emph {et~al.}(1997)\citenamefont
  {Chetyrkin}, \citenamefont {Misiak},\ and\ \citenamefont {Munz}}]{Misiak}%
  \BibitemOpen
  \bibfield  {author} {\bibinfo {author} {\bibfnamefont {K.~G.}\ \bibnamefont
  {Chetyrkin}}, \bibinfo {author} {\bibfnamefont {M.}~\bibnamefont {Misiak}}, \
  and\ \bibinfo {author} {\bibfnamefont {M.}~\bibnamefont {Munz}},\ }\href
  {\doibase 10.1016/S0370-2693(97)00324-9} {\bibfield  {journal} {\bibinfo
  {journal} {Phys. Lett.}\ }\textbf {\bibinfo {volume} {B400}},\ \bibinfo
  {pages} {206} (\bibinfo {year} {1997})},\ \bibinfo {note} {[Erratum: Phys.
  Lett.B425,414(1998)]},\ \Eprint {http://arxiv.org/abs/hep-ph/9612313}
  {arXiv:hep-ph/9612313 [hep-ph]} \BibitemShut {NoStop}%
%%CITATION = HEP-PH/9612313;%%
\bibitem [{\citenamefont {Hewett}\ and\ \citenamefont
  {Rizzo}(1987)}]{HewettRizzo}%
  \BibitemOpen
  \bibfield  {author} {\bibinfo {author} {\bibfnamefont {J.~L.}\ \bibnamefont
  {Hewett}}\ and\ \bibinfo {author} {\bibfnamefont {T.~G.}\ \bibnamefont
  {Rizzo}},\ }\href {\doibase 10.1103/PhysRevD.35.2194} {\bibfield  {journal}
  {\bibinfo  {journal} {Phys. Rev.}\ }\textbf {\bibinfo {volume} {D35}},\
  \bibinfo {pages} {2194} (\bibinfo {year} {1987})}\BibitemShut {NoStop}%
%%CITATION = PHRVA,D35,2194;%%
\bibitem [{\citenamefont {Alok}\ \emph {et~al.}(2015)\citenamefont {Alok},
  \citenamefont {Banerjee}, \citenamefont {Kumar}, \citenamefont {Sankar},\
  and\ \citenamefont {London}}]{Alok_Uma}%
  \BibitemOpen
  \bibfield  {author} {\bibinfo {author} {\bibfnamefont {A.~K.}\ \bibnamefont
  {Alok}}, \bibinfo {author} {\bibfnamefont {S.}~\bibnamefont {Banerjee}},
  \bibinfo {author} {\bibfnamefont {D.}~\bibnamefont {Kumar}}, \bibinfo
  {author} {\bibfnamefont {S.~U.}\ \bibnamefont {Sankar}}, \ and\ \bibinfo
  {author} {\bibfnamefont {D.}~\bibnamefont {London}},\ }\href {\doibase
  10.1103/PhysRevD.92.013002} {\bibfield  {journal} {\bibinfo  {journal} {Phys.
  Rev.}\ }\textbf {\bibinfo {volume} {D92}},\ \bibinfo {pages} {013002}
  (\bibinfo {year} {2015})},\ \Eprint {http://arxiv.org/abs/1504.00517}
  {arXiv:1504.00517 [hep-ph]} \BibitemShut {NoStop}%
%%CITATION = ARXIV:1504.00517;%%
\bibitem [{\citenamefont {Pati}\ and\ \citenamefont
  {Salam}(1974)}]{LR_model_1}%
  \BibitemOpen
  \bibfield  {author} {\bibinfo {author} {\bibfnamefont {J.~C.}\ \bibnamefont
  {Pati}}\ and\ \bibinfo {author} {\bibfnamefont {A.}~\bibnamefont {Salam}},\
  }\href {\doibase 10.1103/PhysRevD.10.275, 10.1103/PhysRevD.11.703.2}
  {\bibfield  {journal} {\bibinfo  {journal} {Phys. Rev.}\ }\textbf {\bibinfo
  {volume} {D10}},\ \bibinfo {pages} {275} (\bibinfo {year} {1974})},\ \bibinfo
  {note} {[Erratum: Phys. Rev.D11,703(1975)]}\BibitemShut {NoStop}%
%%CITATION = PHRVA,D10,275;%%
\bibitem [{\citenamefont {Mohapatra}\ and\ \citenamefont
  {Pati}(1975)}]{LR_model_2}%
  \BibitemOpen
  \bibfield  {author} {\bibinfo {author} {\bibfnamefont {R.~N.}\ \bibnamefont
  {Mohapatra}}\ and\ \bibinfo {author} {\bibfnamefont {J.~C.}\ \bibnamefont
  {Pati}},\ }\href {\doibase 10.1103/PhysRevD.11.2558} {\bibfield  {journal}
  {\bibinfo  {journal} {Phys. Rev.}\ }\textbf {\bibinfo {volume} {D11}},\
  \bibinfo {pages} {2558} (\bibinfo {year} {1975})}\BibitemShut {NoStop}%
%%CITATION = PHRVA,D11,2558;%%
\bibitem [{\citenamefont {Senjanovic}\ and\ \citenamefont
  {Mohapatra}(1975)}]{LR_model_3}%
  \BibitemOpen
  \bibfield  {author} {\bibinfo {author} {\bibfnamefont {G.}~\bibnamefont
  {Senjanovic}}\ and\ \bibinfo {author} {\bibfnamefont {R.~N.}\ \bibnamefont
  {Mohapatra}},\ }\href {\doibase 10.1103/PhysRevD.12.1502} {\bibfield
  {journal} {\bibinfo  {journal} {Phys. Rev.}\ }\textbf {\bibinfo {volume}
  {D12}},\ \bibinfo {pages} {1502} (\bibinfo {year} {1975})}\BibitemShut
  {NoStop}%
%%CITATION = PHRVA,D12,1502;%%
\bibitem [{\citenamefont {Aad}\ \emph {et~al.}(2012)\citenamefont {Aad} \emph
  {et~al.}}]{G.Aad}%
  \BibitemOpen
  \bibfield  {author} {\bibinfo {author} {\bibfnamefont {G.}~\bibnamefont
  {Aad}} \emph {et~al.} (\bibinfo {collaboration} {ATLAS}),\ }\href {\doibase
  10.1140/epjc/s10052-012-2241-5} {\bibfield  {journal} {\bibinfo  {journal}
  {Eur. Phys. J.}\ }\textbf {\bibinfo {volume} {C72}},\ \bibinfo {pages} {2241}
  (\bibinfo {year} {2012})},\ \Eprint {http://arxiv.org/abs/1209.4446}
  {arXiv:1209.4446 [hep-ex]} \BibitemShut {NoStop}%
%%CITATION = ARXIV:1209.4446;%%
\bibitem [{\citenamefont {Chatrchyan}\ \emph {et~al.}(2012)\citenamefont
  {Chatrchyan} \emph {et~al.}}]{Chatrchyan}%
  \BibitemOpen
  \bibfield  {author} {\bibinfo {author} {\bibfnamefont {S.}~\bibnamefont
  {Chatrchyan}} \emph {et~al.} (\bibinfo {collaboration} {CMS}),\ }\href
  {\doibase 10.1007/JHEP08(2012)023} {\bibfield  {journal} {\bibinfo  {journal}
  {JHEP}\ }\textbf {\bibinfo {volume} {08}},\ \bibinfo {pages} {023} (\bibinfo
  {year} {2012})},\ \Eprint {http://arxiv.org/abs/1204.4764} {arXiv:1204.4764
  [hep-ex]} \BibitemShut {NoStop}%
%%CITATION = ARXIV:1204.4764;%%
\bibitem [{\citenamefont {Cho}\ and\ \citenamefont {Misiak}(1994)}]{Cho}%
  \BibitemOpen
  \bibfield  {author} {\bibinfo {author} {\bibfnamefont {P.~L.}\ \bibnamefont
  {Cho}}\ and\ \bibinfo {author} {\bibfnamefont {M.}~\bibnamefont {Misiak}},\
  }\href {\doibase 10.1103/PhysRevD.49.5894} {\bibfield  {journal} {\bibinfo
  {journal} {Phys. Rev.}\ }\textbf {\bibinfo {volume} {D49}},\ \bibinfo {pages}
  {5894} (\bibinfo {year} {1994})},\ \Eprint
  {http://arxiv.org/abs/hep-ph/9310332} {arXiv:hep-ph/9310332 [hep-ph]}
  \BibitemShut {NoStop}%
%%CITATION = HEP-PH/9310332;%%
\bibitem [{\citenamefont {Kou}\ \emph {et~al.}(2013)\citenamefont {Kou},
  \citenamefont {Lü},\ and\ \citenamefont {Yu}}]{Kou}%
  \BibitemOpen
  \bibfield  {author} {\bibinfo {author} {\bibfnamefont {E.}~\bibnamefont
  {Kou}}, \bibinfo {author} {\bibfnamefont {C.-D.}\ \bibnamefont {Lü}}, \ and\
  \bibinfo {author} {\bibfnamefont {F.-S.}\ \bibnamefont {Yu}},\ }\href
  {\doibase 10.1007/JHEP12(2013)102} {\bibfield  {journal} {\bibinfo  {journal}
  {JHEP}\ }\textbf {\bibinfo {volume} {12}},\ \bibinfo {pages} {102} (\bibinfo
  {year} {2013})},\ \Eprint {http://arxiv.org/abs/1305.3173} {arXiv:1305.3173
  [hep-ph]} \BibitemShut {NoStop}%
%%CITATION = ARXIV:1305.3173;%%
\bibitem[]{to_come}
   Under preparation (S.Mandal, A.Biswas, N.Sinha).
\bibitem [{\citenamefont {Langacker}\ and\ \citenamefont
  {Sankar}(1989)}]{Langacker}%
  \BibitemOpen
  \bibfield  {author} {\bibinfo {author} {\bibfnamefont {P.}~\bibnamefont
  {Langacker}}\ and\ \bibinfo {author} {\bibfnamefont {S.~U.}\ \bibnamefont
  {Sankar}},\ }\href {\doibase 10.1103/PhysRevD.40.1569} {\bibfield  {journal}
  {\bibinfo  {journal} {Phys. Rev.}\ }\textbf {\bibinfo {volume} {D40}},\
  \bibinfo {pages} {1569} (\bibinfo {year} {1989})}\BibitemShut {NoStop}%
%%CITATION = PHRVA,D40,1569;%%
\bibitem [{\citenamefont {Abdesselam}\ \emph {et~al.}(2016)\citenamefont
  {Abdesselam} \emph {et~al.}}]{Belle_rad}%
  \BibitemOpen
  \bibfield  {author} {\bibinfo {author} {\bibfnamefont {A.}~\bibnamefont
  {Abdesselam}} \emph {et~al.} (\bibinfo {collaboration} {Belle}),\ }\bibfield
  {booktitle} {\emph {\bibinfo {booktitle} {{30th Rencontres de Physique de La
  Vallée d'Aoste La Thuile, Aosta valley, Italy, March 6-12, 2016}}},\ }\href
  {\doibase 10.1103/PhysRevLett.118.051801} {\  (\bibinfo {year} {2016}),\
  10.1103/PhysRevLett.118.051801},\ \bibinfo {note} {[Phys. Rev.
  Lett.118,051801(2017)]},\ \Eprint {http://arxiv.org/abs/1603.03257}
  {arXiv:1603.03257 [hep-ex]} \BibitemShut {NoStop}%
%%CITATION = ARXIV:1603.03257;%%
\bibitem [{\citenamefont {Patrignani}(2016)}]{pdg}%
  \BibitemOpen
  \bibfield  {author} {\bibinfo {author} {\bibfnamefont {C.}~\bibnamefont
  {Patrignani}},\ }\href {\doibase 10.1088/1674-1137/40/10/100001} {\bibfield
  {journal} {\bibinfo  {journal} {Chin. Phys.}\ }\textbf {\bibinfo {volume}
  {C40}},\ \bibinfo {pages} {100001} (\bibinfo {year} {2016})}\BibitemShut
  {NoStop}%
%%CITATION = CHPHD,C40,100001;%%
\bibitem [{\citenamefont {Aaij}\ \emph {et~al.}(2014)\citenamefont {Aaij} \emph
  {et~al.}}]{LHCb}%
  \BibitemOpen
  \bibfield  {author} {\bibinfo {author} {\bibfnamefont {R.}~\bibnamefont
  {Aaij}} \emph {et~al.} (\bibinfo {collaboration} {LHCb}),\ }\href {\doibase
  10.1103/PhysRevLett.112.161801} {\bibfield  {journal} {\bibinfo  {journal}
  {Phys. Rev. Lett.}\ }\textbf {\bibinfo {volume} {112}},\ \bibinfo {pages}
  {161801} (\bibinfo {year} {2014})},\ \Eprint {http://arxiv.org/abs/1402.6852}
  {arXiv:1402.6852 [hep-ex]} \BibitemShut {NoStop}%
%%CITATION = ARXIV:1402.6852;%%
\bibitem [{\citenamefont {de~Boer}\ and\ \citenamefont
  {Hiller}(2017)}]{Hiller}%
  \BibitemOpen
  \bibfield  {author} {\bibinfo {author} {\bibfnamefont {S.}~\bibnamefont
  {de~Boer}}\ and\ \bibinfo {author} {\bibfnamefont {G.}~\bibnamefont
  {Hiller}},\ }\href@noop {} {\  (\bibinfo {year} {2017})},\ \Eprint
  {http://arxiv.org/abs/1701.06392} {arXiv:1701.06392 [hep-ph]} \BibitemShut
  {NoStop}%
%%CITATION = ARXIV:1701.06392;%%
\bibitem [{\citenamefont {Grossman}\ and\ \citenamefont
  {Pirjol}(2000)}]{Grossman}%
  \BibitemOpen
  \bibfield  {author} {\bibinfo {author} {\bibfnamefont {Y.}~\bibnamefont
  {Grossman}}\ and\ \bibinfo {author} {\bibfnamefont {D.}~\bibnamefont
  {Pirjol}},\ }\href {\doibase 10.1088/1126-6708/2000/06/029} {\bibfield
  {journal} {\bibinfo  {journal} {JHEP}\ }\textbf {\bibinfo {volume} {06}},\
  \bibinfo {pages} {029} (\bibinfo {year} {2000})},\ \Eprint
  {http://arxiv.org/abs/hep-ph/0005069} {arXiv:hep-ph/0005069 [hep-ph]}
  \BibitemShut {NoStop}%
%%CITATION = HEP-PH/0005069;%%
\bibitem [{\citenamefont {Fajfer}\ and\ \citenamefont
  {Kamenik}(2005)}]{fajfer_ff}%
  \BibitemOpen
  \bibfield  {author} {\bibinfo {author} {\bibfnamefont {S.}~\bibnamefont
  {Fajfer}}\ and\ \bibinfo {author} {\bibfnamefont {J.}~\bibnamefont
  {Kamenik}},\ }\href {\doibase 10.1103/PhysRevD.72.034029} {\bibfield
  {journal} {\bibinfo  {journal} {Phys. Rev. D}\ }\textbf {\bibinfo {volume}
  {72}},\ \bibinfo {pages} {034029} (\bibinfo {year} {2005})}\BibitemShut
  {NoStop}%
\bibitem [{\citenamefont {Cao}(2012)}]{eta}%
  \BibitemOpen
  \bibfield  {author} {\bibinfo {author} {\bibfnamefont {F.-G.}\ \bibnamefont
  {Cao}},\ }\href {\doibase 10.1103/PhysRevD.85.057501} {\bibfield  {journal}
  {\bibinfo  {journal} {Phys. Rev.}\ }\textbf {\bibinfo {volume} {D85}},\
  \bibinfo {pages} {057501} (\bibinfo {year} {2012})},\ \Eprint
  {http://arxiv.org/abs/1202.6075} {arXiv:1202.6075 [hep-ph]} \BibitemShut
  {NoStop}%
%%CITATION = ARXIV:1202.6075;%%
\bibitem [{\citenamefont {Bediaga}\ \emph {et~al.}(2004)\citenamefont
  {Bediaga}, \citenamefont {Navarra},\ and\ \citenamefont {Nielsen}}]{f0980}%
  \BibitemOpen
  \bibfield  {author} {\bibinfo {author} {\bibfnamefont {I.}~\bibnamefont
  {Bediaga}}, \bibinfo {author} {\bibfnamefont {F.~S.}\ \bibnamefont
  {Navarra}}, \ and\ \bibinfo {author} {\bibfnamefont {M.}~\bibnamefont
  {Nielsen}},\ }\href {\doibase
  http://dx.doi.org/10.1016/j.physletb.2003.10.102} {\bibfield  {journal}
  {\bibinfo  {journal} {Physics Letters B}\ }\textbf {\bibinfo {volume}
  {579}},\ \bibinfo {pages} {59 } (\bibinfo {year} {2004})}\BibitemShut
  {NoStop}%
\bibitem [{\citenamefont {Aydin}\ and\ \citenamefont {Yilmaz}(2006)}]{a0980}%
  \BibitemOpen
  \bibfield  {author} {\bibinfo {author} {\bibfnamefont {C.}~\bibnamefont
  {Aydin}}\ and\ \bibinfo {author} {\bibfnamefont {A.~H.}\ \bibnamefont
  {Yilmaz}},\ }\href {\doibase 10.1142/S0217732306019955} {\bibfield  {journal}
  {\bibinfo  {journal} {Mod. Phys. Lett.}\ }\textbf {\bibinfo {volume} {A21}},\
  \bibinfo {pages} {1297} (\bibinfo {year} {2006})},\ \Eprint
  {http://arxiv.org/abs/hep-ph/0503128} {arXiv:hep-ph/0503128 [hep-ph]}
  \BibitemShut {NoStop}%
%%CITATION = HEP-PH/0503128;%%
\bibitem [{\citenamefont {Ebert}\ \emph {et~al.}(2006)\citenamefont {Ebert},
  \citenamefont {Faustov},\ and\ \citenamefont {Galkin}}]{vector_f}%
  \BibitemOpen
  \bibfield  {author} {\bibinfo {author} {\bibfnamefont {D.}~\bibnamefont
  {Ebert}}, \bibinfo {author} {\bibfnamefont {R.~N.}\ \bibnamefont {Faustov}},
  \ and\ \bibinfo {author} {\bibfnamefont {V.~O.}\ \bibnamefont {Galkin}},\
  }\href {\doibase 10.1016/j.physletb.2006.02.042} {\bibfield  {journal}
  {\bibinfo  {journal} {Phys. Lett.}\ }\textbf {\bibinfo {volume} {B635}},\
  \bibinfo {pages} {93} (\bibinfo {year} {2006})},\ \Eprint
  {http://arxiv.org/abs/hep-ph/0602110} {arXiv:hep-ph/0602110 [hep-ph]}
  \BibitemShut {NoStop}%
%%CITATION = HEP-PH/0602110;%%
\bibitem [{\citenamefont {Narison}(2016)}]{Dst_f}%
  \BibitemOpen
  \bibfield  {author} {\bibinfo {author} {\bibfnamefont {S.}~\bibnamefont
  {Narison}},\ }\bibfield  {booktitle} {\emph {\bibinfo {booktitle}
  {{Proceedings, 18th High-Energy Physics International Conference in Quantum
  Chromodynamics (QCD 15): Montpellier, France, June 29-July 03, 2015}}},\
  }\href {\doibase 10.1016/j.nuclphysbps.2016.02.030} {\bibfield  {journal}
  {\bibinfo  {journal} {Nucl. Part. Phys. Proc.}\ }\textbf {\bibinfo {volume}
  {270-272}},\ \bibinfo {pages} {143} (\bibinfo {year} {2016})},\ \Eprint
  {http://arxiv.org/abs/1511.05903} {arXiv:1511.05903 [hep-ph]} \BibitemShut
  {NoStop}%
%%CITATION = ARXIV:1511.05903;%%
\end{thebibliography}

\end{document}